\documentclass[12pt]{article}

\usepackage{fullpage}
\usepackage{amsthm}
\usepackage{amsfonts}
\usepackage{amsmath}
\usepackage{amssymb}
\usepackage{array}
\usepackage{bm}
\usepackage{enumerate}
\usepackage{isomath}
\usepackage[labelfont=bf]{caption}
\usepackage{subcaption}
\usepackage{tabularx}
    \newcolumntype{V}{>{\centering\arraybackslash}X}
    \newcolumntype{B}[1]{>{\centering\arraybackslash}m{#1}}
\usepackage{tabularray}
\usepackage{tikz}
\usepackage{graphicx}
\usepackage[hang,flushmargin]{footmisc} 
\usepackage{framed}
\usepackage[normalem]{ulem}
\usepackage{verbatim}
\usepackage{soul}

\newcolumntype{S}{>{\centering\arraybackslash}p{3.5em}}
\newcolumntype{M}{>{\centering\arraybackslash}p{3.5em}}
\newcolumntype{C}{>{\centering\arraybackslash}p{6em}}
\newcolumntype{G}{>{\centering\arraybackslash}p{8em}}
\newcolumntype{J}{>{\centering\arraybackslash}p{12em}}
\newcolumntype{L}{>{\centering\arraybackslash}p{16em}}
\usepackage{natbib}
\numberwithin{equation}{section}
\usepackage{sectsty}
\allsectionsfont{\color{violet}}  
\usepackage{xcolor}

\newcommand{\bx}{\mathbf{x}}
\newcommand{\bR}{\mathbf{R}}
\newcommand{\by}{\mathbf{y}}
\newcommand{\D}{\mathcal{D}}
\newcommand{\MVN}{\mathsf{MVN}}
\newcommand{\bbeta}{\boldsymbol{\beta}}
\newcommand{\bmu}{\boldsymbol{\mu}}
\newcommand{\blambda}{\boldsymbol{\lambda}}
\newcommand{\hx}{\mathbf{h}(\mathbf{x})}
\newcommand{\hxD}{\mathbf{h}(\mathbf{x}^\mathcal{D})}
\newcommand{\hxstar}{\mathbf{h}(\mathbf{x}^*)}

\newcommand{\htxD}{\mathbf{h}^\top(\mathbf{x}^\mathcal{D})}

\title{\color{violet} {\large \textbf{VPPE: Application of Scaled Vecchia Approximations to Parallel Partial Emulation}}}
\author{\color{violet} {\large Josh Seidman\thanks{Department of Mathematical and Statistical Sciences, Marquette University, Milwaukee, WI 53201 USA},
Elaine T. Spiller\footnotemark[1]}}
\date{}

\begin{document}
\maketitle
\section{Abstract}
Computer models or simulators are widely used across scientific fields, but are computationally expensive limiting their use to explore possible scenarios/outcomes. Gaussian process emulators are statistical surrogates that can rapidly approximate the outputs of computer models at untested inputs and enable uncertainty quantification studies. The parallel partial emulation (PPE) was developed to model simulators with vector-valued outputs. While the PPE is adept at fitting simulator data with multidimensional outputs, the time to fit the PPE increases quickly as the number of training runs increases. The Scaled Vecchia approximation, a fast approximation to multivariate Gaussian likelihoods, makes fitting Gaussian process emulators with large training datasets  tractable. Here we introduce the Vecchia Parallel Partial Emulation (VPPE) that utilizes the Scaled Vecchia approximation within the PPE framework to allow for parallel partial emulation with larger training datasets. The VPPE is applied to three computer experiments, a synthetic data set, a hydrology model, and a volcanic flow model, yielding comparable predictive accuracy to the PPE at a fraction of the runtime.
\section{Introduction}

Uncertainty quantification techniques rely heavily on surrogates of complex computer models, or {simulators}, of physical, engineering, and biological systems. Since its introduction, Gaussian process-based (GP) emulation \citep{Sack:Schi:Welc:1989, Curr:etal:1988, Welch:1992} has adapted markedly to address challenges posed by data from more and more complicated simulators \citep{Sant:Will:Notz:2018, Gram:2020}. Here we seek to address three challenges -- large-dimensional output, large training data sets, and the complicated likelihood surfaces involved in fitting GPs -- by developing the {\it Vecchia Parallel Partial Emulator}.

Very high dimensional outputs are common in geo-spatial models where the quantity of interest -- snow depth on ice \citep{Lawrence:etal:2024}, depth of flooding inundation \citep{Wijaya:etal:2023}, infectious disease levels \citep{Balcan:etal:2010}, etc.~-- varies in space resulting in output dimensions of $10^4-10^7$ per simulation. Parallel partial emulation (PPE) \citep{Gu:Berg:2016} is an appealing ``global" emulation option as it inherits conservation properties (mass, momentum, etc.) of the simulator \citep{Gao:Pit:2024} and is computationally tractable for a modest number of training simulations. Most ``local" GP methods treat space as input \citep{Gram:Aple:2015, Gramacy:2016} and can be quite effective, but become computationally burdensome when the number of spatial components is very large. The Vecchia approximation \citep{Vecchia:1988} applied to GP emulation is a rapidly advancing tool that handles a large number of training simulations quite efficiently \citep{Katz:Guin:Lawr:2022}. Here we exploit the structure of the Scaled Vecchia likelihood approximation for GPs in modeling high output-dimensional simulator data with PPEs in the case of  moderate to large numbers of training simulations.

Rather than relying on MLEs for Vecchia likelihood approximations of GP range parameters, we adopt the Robust GaSP framework proposed by  \cite{Gu:Wang:Berg:2018, Gu:Palo:Berg:2019}. In Section~\ref{sec:Methods}, we first derive the {\it Vecchia marginal posterior}. Marginalizing out the GP trend and variance parameters leads to a ``robust" (as defined in \cite{Gu:Wang:Berg:2018}) posterior probability distribution for finding optimal GP range parameters. Then we derive the gradient of the Vecchia marginal posterior, and finally the {\it Vecchia Parallel Partial Emulator}.

To demonstrate the VPPE's efficacy and speed, we apply it to three increasingly realistic large output-dimensional simulators: a synthetic GP simulator, a ground water flow simulator, and a volcanic flow simulator. In all cases the VPPE and PPE are nearly identically in terms of predictive performance, but for large training data sets, fitting a VPPE requires only a small fraction of the time required to fit a PPE. Details of these experiments and results are in Section~\ref{sec:Results}. We start with background on GPs, the Vecchia approximation, Robust GaSP, PPEs in Section~\ref{sec:Background} and conclude with a discussion in Section~\ref{sec:Conclusion}.

\section{Background}\label{sec:Background}
\subsection{Gaussian process emulation \label{sec:GP}}
Gaussian process emulation is frequently used as a surrogate for computer simulators that are computationally intensive to run \citep{Sack:Schi:Welc:1989, Curr:etal:1988, Welch:1992}. Let $\bx \in \mathcal{X}$ be a $p$-dimensional vector of inputs, $\bx = (x_1,\dots,x_p)^\top$,  such that $\mathbf{y(x)}$ is the simulator's $k$-dimensional real-valued output. In the scalar-output case ($k=1$), one can model $y$ via a Gaussian process,
\begin{align}
    y(\cdot) \sim \mathcal{GP}\big{(}\mu(\cdot),C(\cdot,\cdot)\big{)}, \label{eq:GP}
\end{align}
with mean trend function, $\mu(\cdot)$, covariance function, $C(\cdot,\cdot) = \sigma^2 c(\cdot,\cdot)$, unknown variance, $\sigma^2$, and correlation function, $c(\cdot,\cdot)$. 

Consider a set of $n$ input configurations, or {\it design}, $\bx^\D =\{\bx_1,\dots,\bx_n\}$ where $\bx_i = (x_{i1},\dots,x_{ip})^\top$ and the corresponding computer model responses, $\by^\D =\big{(}y(\bx_1), \dots, y(\bx_n)\big{)}^
\top$. Together this design/response pair can be modeled as a multivariate Normal, 
\begin{align}\label{eq:MVN}
\by^\D \mid \bbeta, \sigma^2, \blambda \sim \MVN\big{(}\bmu,\sigma^2\mathbf{R} \big{)}, 
\end{align}
where $\bbeta$ and $\blambda$ are unknown trend and range parameters, respectively.  
The mean trend function is taken to have the form,
$\mu(\bx) = \mathbf{h(x)} \boldsymbol{\beta}$,
where $\mu$ is linear in terms of the covariate parameters $\boldsymbol{\beta}$. Further, $\bmu$ is the $n$-vector of the trend evaluated at each design point in $\bx^\D$. Common assumptions for the mean trend's basis functions include $\mathbf{h(x)} = 1$ or $\mathbf{h(x)} = (1,\bx^\top)$, where $\mathbf{h}(\bx)$ is a $q$-dimensional vector. The $(i,j)$ element of the correlation matrix, $\mathbf{R}$, is $c(\bx_i, \bx_j)$  \citep{Will:Rasm:Edwa:2006}.  Here $c(\bx_i, \bx_j)$ is taken to be the product correlation across all input dimensions \citep{Sant:Will:2003},
\begin{align}
    c(\bx_i, \bx_j) = \prod_{l=1}^p c_l(x_{il}, x_{jl}; \lambda_l), \label{eq:corr_fun}
\end{align}
where $c_l(\cdot,\cdot)$ is the correlation function for the $l^{th}$ input dimension, and $\lambda_l$ is the range parameter for the $l^{th}$ input dimension. Common correlation functions used in GP emulation for computer models include the Matérn family \citep{Stein:1999} and the exponential family \citep{Will:Rasm:Edwa:2006}. The correlation functions that we consider in this paper are the power exponential functions, the Matérn 3/2, and the Matérn 5/2 (Table \ref{tab:corr_functions}).

The likelihood function associated with Eq.~\ref{eq:MVN} has the form,
\begin{align}
    {L}(\bbeta, \sigma^2, \blambda \mid \bx^\D, \by^\D) &= (2\pi)^{-n/2} |\sigma^2 \mathbf{R}|^{-1/2} \exp \left( -\frac{1}{2} (\by^\D - \bmu)^\top (\sigma^2 \mathbf{R})^{-1} (\by^\D - \bmu) \right). \label{eq:likelihood}
\end{align}
Estimating the range parameters, $\blambda$, is the key task in ``fitting" a GP emulator. This estimation requires the repeated evaluation of the likelihood function, which can be computationally expensive when $n$ is large. The Robust GaSP (RGaSP) approach \citep{Gu:Wang:Berg:2018} to fitting GPs and its associated package, \texttt{RobustGaSP}, consider a posterior distribution for $\blambda$ with an integrated likelihood (over $\bbeta$ and $\sigma^2$) and with a prior on the range parameters. Mode posterior estimates then yield optimal range parameters. Using the Scaled Vecchia approximation \citep{Katz:Guin:Lawr:2022} to approximate the integrated likelihood function will allow for a much quicker evaluation when the number of design points is large.

\begin{table}[]
\centering
\begin{tabular}{ll}
\hline
Power exponential  & $\exp\left( -\left(\frac{d}{\lambda}\right)^\alpha \right), \alpha \in [1,2]$  \\
Matérn 3/2 & $\left(1+\frac{\sqrt{3}d}{\lambda}\right)\exp\left(-\frac{\sqrt{3}d}{\lambda}\right)$  \\
Matérn 5/2 & $\left(1+\frac{\sqrt{5}d}{\lambda} + \frac{5d^2}{3\lambda^2}\right)\exp\left(-\frac{\sqrt{5}d}{\lambda}\right)$ \\
\hline
\end{tabular}
\caption{The correlation functions implemented in \texttt{RobustGaSPV} where $d$ is the distance between two points in a dimension, and $\lambda$ is the range parameter for that dimension.}
\label{tab:corr_functions}
\end{table}

\subsection{Vecchia approximations \label{sec:Vecchia}}
Vecchia's approximation \citep{Vecchia:1988} is a useful tool from spatial statistics to approximate likelihoods that are expensive to evaluate. 
Consider a model $y = \mathbf{h}(\bx)\boldsymbol{\beta} + \xi + \eta$, where $\xi$ is a zero mean Gaussian process and $\mathbf{h}(\bx)\boldsymbol{\beta}$ is a user-specified mean trend. Further, the observations have measurement errors, $\eta_i \overset{\text{i.i.d.}}{\sim} \mathcal{N}(0,\sigma^2_{\eta})$.
Considering $n$ input configurations, the joint density of the GP model's outputs, $\mathbf{y}$, can be written as $p(\mathbf{y}) = \prod_{i=1}^n p(y_i | y_1,\dots,y_{i-1})$. Vecchia \citep{Vecchia:1988} gives the following approximation,
\begin{align}
    \hat{p}(\mathbf{y})  = \prod_{i=1}^n p\left(y_i|\mathbf{y}_{b(i)}\right), \label{eq:p_hat}
\end{align}
where $b(i) \subset \{1,\dots,i-1\}$ is a conditioning set of size $m$ (or more precisely, $\min(m,i-1)$) for each $i=2,\dots,n$, and $b(1) = \emptyset$. Even when the size of the conditioning set is small, ($m \ll n$), this approximation can be extremely accurate due to the screening effect \citep{Stein:2002}. Each $p(y_i|\mathbf{y}_{b(i)})$ is a Gaussian distribution and can be computed in parallel with the computational complexity of $\mathcal{O}(m^3)$, leading to an overall computational complexity of $\mathcal{O}(nm^3)$ for the evaluation of the approximate distribution $\hat{p}(\mathbf{y})$. 

If one conditions on $m$ nearest neighbors, the approximate distribution in Eq. \ref{eq:p_hat} can be used to approximate the likelihood function as shown in \cite{Vecchia:1988}. This Vecchia likelihood is given by,
\begin{align}
    {L}_m(\blambda, \bbeta, \sigma^2 \mid \bx^\D, \by^\D) = (2\pi)^{-n/2}(\sigma^2)^{-n/2}\left( \prod_{i=1}^n \omega_{im}^{-1/2} \right) \exp\left[ -(2\sigma^2)^{-1} \sum_{i=1}^n \omega_{im}^{-1} e_{im}^2 \right]. \label{eq:Vecchia_lik}
\end{align}
The terms in Eq.~\ref{eq:Vecchia_lik} are given by,
\begin{align}
    \omega_{im} &= 1 + \nu^2 - \mathbf{r}^\top_{im}(\bR_{im} + \nu^2 I)^{-1}\mathbf{r}_{im}, \label{eq:vec_cond_var}\\
    e_{im} &= \varepsilon_i - \mathbf{r}^\top_{im}(\bR_{im} + \nu^2 I)^{-1} \boldsymbol{\varepsilon}_{im}, \label{eq:vec_cond_residual}\\
    \mathbf{r}_{im} &= c(\bx_i,\bx_{im}; \boldsymbol{\lambda}) ,\label{eq:vec_corr_i}\\
    \bR_{im} &= c(\bx_{im},\bx_{im}; \boldsymbol{\lambda}), \label{eq:vec_corr_im}
\end{align}
where $\bx_i$ is the $i^{th}$ input data point in $\bx^{\D}$, and $\bx_{im}$ is the set of $m$ nearest neighbors to $\bx_i$. The term $\nu^2$ is the nugget-variance ratio $\sigma^2_{\eta}/\sigma^2$, $\varepsilon_i = y_i - \mathbf{h}(x_i)\beta$, and $\varepsilon_{im} = y_{im} - \mathbf{h}(\bx_{im})\beta$. $\mathbf{r}_{im}$ is the $m \times 1$ correlation vector between $\bx_i$ and $\bx_{im}$, and $\bR_{im}$ is the $m \times m$ correlation matrix of $\bx_{im}$. Note, for the $i^{\text{th}}$ data point, Eq. \ref{eq:vec_cond_var} is analogous to the predictive variance conditioned on the nearest $m$ neighbors, and Eq. \ref{eq:vec_cond_residual} is analogous to the conditional residual \citep{Sant:Will:Notz:2018}.

The joint distribution $\hat{p}(\mathbf{y})$ is multivariate Gaussian and for $m=n-1$, the approximation becomes exact such that $\hat{p}(\mathbf{y}) = p(\mathbf{y})$. Unlike local GP approximations such as \verb|laGP| \citep{Gramacy:2016}, the Vecchia approximation can also be use to estimate model parameters and make global predictions. Vecchia's flexibility to move between local and global modeling proves particularly useful in cases of large-training data sets and large output dimensions as we will see in Section~\ref{sec:Results}.

The Vecchia approximation depends on the order $y_1,\dots,y_n$ and of the conditioning set $b(i)$. A common implementation uses a maximum-minimum distance (maximin) ordering and nearest-neighbor (NN) conditioning \citep{Guinness:2018}. Maximin chooses the first point randomly then proceeds by choosing each following point that maximizes the minimum distance to the previous points in the ordering. This method effectively chooses subsequent points that would maximize the amount of new information encompassed by the previous points and the new one.

\subsubsection{Scaled Vecchia approximations \label{sec:SVecchia}}
Since different input variables can have varying effects on the response variables, the standard maximin ordering and NN conditioning can lead to ineffective choices for the nearest neighbors, resulting in a poor approximation. Take initial range parameters to be $\boldsymbol{\lambda} = (\lambda_1, \dots, \lambda_p)^\top$. In Scaled Vecchia \citep{Katz:Guin:Lawr:2022}, the initial range parameter for a dimension are chosen to be a fifth of the range of the input data in that dimension. Instead of determining the ordering and conditioning in the standard input space $\mathcal{X}$, one can do so in the scaled input space $\tilde{\mathcal{X}}$, where a point $\bx = (x_1,\dots,x_p)^\top$ in $\mathcal{X}$ becomes $\tilde{\bx} = (x_1/\lambda_1,\dots,x_p/\lambda_p)^\top$ in $\tilde{\mathcal{X}}$. Scaled Vecchia approximation \citep{Katz:Guin:Lawr:2022} deviates from the standard Vecchia approach by performing maximin ordering and NN conditioning on the scaled inputs $\{\mathbf{\tilde{x}_1,\dots,\tilde{x}_n}\}$ instead of the original inputs $\{\mathbf{x_1,\dots,x_n}\}$. It has been shown that the ordering and conditioning can be computed in near linear time in $n$ \citep{Schaf:Sulli:Owha:2021}.

In Scaled Vecchia approximation, the range parameters in the covariance function are estimated by using the Fisher scoring algorithm \citep{Guinness:2019}, which utilizes the profiled likelihood. In Section \ref{sec:Methods}, we replace the Fisher scoring algorithm with maximization of the posterior distribution over the range parameters. The maximin ordering and NN conditioning will be identical to that in Scaled Vecchia approximation.

\subsection{RobustGaSP}
Maximum likelihood estimations (MLEs) of range parameters are common for fitting GP emulators. Yet MLEs present a difficult optimization problem due to the likelihood surface of GP models being notoriously hard to optimize \citep{Kenn:OHag:2001}. Gu, Wang, and Berger propose a ``Robust GaSP" alternative to MLEs that is easier to numerically optimize \citep{Gu:Wang:Berg:2018}. Implemented in the package \texttt{RobustGaSP} \citep{Gu:Palo:Berg:2019}, the Robust GaSP (RGaSP) method utilizes integrated likelihoods over $\sigma^2$ and $\boldsymbol{\beta}$, objective priors for $\boldsymbol{\lambda}$, and mode posterior estimates for $\boldsymbol{\lambda}$. In this section we will review the statistical framework for scalar outputs ($k=1$). Unknown parameters needed to ``fit" a GP emulator are the mean parameters $(\beta_1,\dots,\beta_q)$, a variance parameter $(\sigma^2)$, and range parameters $(\lambda_1,\dots,\lambda_p)$. The objective prior used in RGaSP has the form,
\begin{align}
    \pi(\blambda, \boldsymbol{\beta},\sigma^2) = \pi(\boldsymbol{\beta}, \sigma^2)\pi(\boldsymbol{\lambda}),  \label{eq:RobustGaSPprior}
\end{align}
where $\pi(\boldsymbol{\lambda})$ is a reference prior for the range parameters and $\pi(\boldsymbol{\beta}, \sigma^2) \propto \frac{1}{\sigma^2}$. For given model runs inputs $\bx^{\mathcal{D}}$, outputs $\mathbf{y}^{\mathcal{D}}$, mean trend function $\mathbf{h}(\cdot)$, and correlation matrix of $\mathbf{R}$, integrating ${L}(\boldsymbol{\lambda}, \boldsymbol{\beta},\sigma^2 \mid \bx^\D, \mathbf{y}^\mathcal{D}) \pi(\boldsymbol{\beta}, \sigma^2)$ over $(\boldsymbol{\beta},\sigma^2)$ (\ref{eq:RobustGaSPprior}) yields the marginal likelihood
\begin{align}
    \mathcal{L}(\boldsymbol{\lambda} \mid \bx^\D, \mathbf{y}^\mathcal{D}) \propto |\mathbf{R}|^{-1/2}|\htxD \mathbf{R}^{-1} \hxD|^{-1/2} \left(S^2 \right)^{-\left(\frac{n-q}{2}\right)}. \label{eq:likelihood}
\end{align}
Here, $S^2 = (\mathbf{y}^\mathcal{D})^\top \mathbf{Q} \mathbf{y}^\mathcal{D}$ and $\mathbf{Q} = \mathbf{R}^{-1} - \mathbf{R}^{-1}\hxD\{\htxD \mathbf{R}^{-1} \hxD \}^{-1} \htxD \mathbf{R}^{-1}$. Additionally, $\hxD$ is the $n \times q$ matrix of the trend basis functions evaluated at the input design points.

The \texttt{RobustGaSP} package includes multiple options for prior functions, as seen in Table \ref{tab:priors}~\citep{Gu:Palo:Berg:2019}. The default choice for the prior is the jointly robust prior. A benefit to the jointly robust prior $\pi^{JR}(\cdot)$ is that it approximates the reference prior quite well and it has closed form derivatives unlike the reference prior. This allows for a more efficient computation of the gradient within optimization routines and results in faster overall computation. Additionally, the \texttt{RobustGaSP} package utilizes the gradient-based low-storage quasi-Newton (L-BFGS) optimization method \citep{Nocedal:1980, Liu:Nocedal:1989} implemented in the package \texttt{nloptr} \citep{Ypma:2014}.

\begin{table}[]
\centering
\begin{tabular}{ll}
\hline
$\pi^R(\boldsymbol{\lambda})$ & $|\mathbf{I}^*(\boldsymbol{\lambda})|^{1/2}$  \\
$\pi^{JR}(\boldsymbol{\lambda})$  & $\left(\sum_{l=1}^p C_l \lambda_l^{-1} \right)^a \exp \left(-b\sum_{l=1}^p C_l\lambda_l^{-1} \right)$ \\
\hline
\end{tabular}
\caption{Prior functions for range parameters available in \texttt{RobustGaSP}. $\mathbf{I}^*(\cdot)$ is the expected Fisher information matrix, after $(\boldsymbol{\beta},\sigma^2)$ are integrated out. The default choices for the hyperparameters in $\pi^{JR}$ are $a=0.2, b=n^{-1/p}(a+p)$, and $C_l$ to be the mean of $|x_{il}^\mathcal{D} - x_{jl}^\mathcal{D}|$ for $1 \leq i,j \leq n, i \neq j$.}
\label{tab:priors}
\end{table}

\subsubsection{Prediction}
Once the mode posterior estimate of the range parameters, $\hat{\boldsymbol{\lambda}}$, has been obtained, and $(\boldsymbol{\beta},\sigma^2)$ have been marginalized out, the predictive distribution of the emulator follows a student's $t$-distribution at new input $\bx^*$ \citep{Gu:Wang:Berg:2018}. For deterministic simulators, the GP emulator is given by,
\begin{align}
    y(\bx^*)\mid \bx^\D \mathbf{y}^\mathcal{D}, \hat{\boldsymbol{\lambda}} \sim t_{n-q}\left(\hat{y}(\bx^*), \hat{\sigma}^2 c^{**}\right), \label{eq:pred_dist}
\end{align}
with $n-q$ degrees of freedom. Here,
\begin{align}
    \hat{y}(\bx^*) &= \mathbf{h}(\bx^*) \hat{\boldsymbol{\beta}} + \mathbf{r}^T(\bx^*)\mathbf{R}^{-1}\Big{(}\mathbf{y}^\mathcal{D} - \hxD \hat{\boldsymbol{\beta}}\Big{)}, \label{eq:yhatxstar_mean}\\
    \hat{\sigma}^2 &= (n-q)^{-1}\Big{(}\mathbf{y}^\mathcal{D} - \hxD\hat{\boldsymbol{\beta}}\Big{)}^\top \mathbf{R}^{-1}\Big{(}\mathbf{y}^\mathcal{D} - \hxD\hat{\boldsymbol{\beta}}\Big{)}, \label{eq:sigmahat2}\\
    \begin{split}
        c^{**} &= 1 - \mathbf{r}^\top(\bx^*) \mathbf{R}^{-1} \mathbf{r}(\bx^*) + \left(\hxstar-\htxD \mathbf{R}^{-1} \mathbf{r}(\bx^*)\right)^\top \\
        &\quad \times \left(\htxD\mathbf{R}^{-1} \hxD\right)^{-1}\left(\hxstar-\htxD\mathbf{R}^{-1}\mathbf{r}(\bx^*)\right), 
    \end{split}\label{eq:c**}
\end{align}
with $\hat{\boldsymbol{\beta}} = (\htxD\mathbf{R}^{-1} \hxD)^{-1} \htxD\mathbf{R}^{-1}\mathbf{y}^\mathcal{D}$ (the generalized least squares estimate for $\boldsymbol{\beta}$), and $\mathbf{r}(\mathbf{x^*}) = c(\bx^*, \bx^\mathcal{D})$. Further, $\mathbf{R}$ and $\hxD$ are as defined in Section \ref{sec:GP}. This emulator is an interpolator of the design points $\bx^\mathcal{D}$ (i.e. $\eta = 0$).

For non-deterministic simulators, $\eta$ is taken to be a zero mean normal random variable. To account for this in our GP, one can parameterize a new covariance function and matrix, along with new definitions for the prior functions in Table \ref{tab:priors}. Further details can be found in Gu, Palomo, and Berger's paper about the \texttt{RobustGaSP} package \citep{Gu:Palo:Berg:2019}. 


\subsection{Parallel partial emulation \label{sec:PPE}}
For many simulators, it is common for the output dimension, $k$, to be quite large, leading to considerable computational complexity for GP emulation. Gu and Berger have developed the method parallel partial Gaussian process emulation (PPE). PPE assumes a common correlation structure among inputs, but unique trend regression coefficients, and scalar variances among output components. A benefit of PPE is that it inherits the smoothness of the simulator as well as some of the underlying physical properties that the simulator obeys \citep{Gu:Berg:2016,Gao:Pit:2024}.

Let $y_j(\bx)$ be the $j$th component of the output of the simulator such that $\mathbf{y}(\bx) = \left(y_1(\bx),\dots,y_k(\bx)\right)^\top$ is the output of the simulator exercised at input $\bx$. $Y^\mathcal{D}$ is the $n \times k$ matrix of simulator outputs given design input values, $\bx^\mathcal{D}$. The PPE assumes independent GPs of the form (\ref{eq:GP}) with mean trends, $\hx\boldsymbol{\beta}_j$, where $\hx$ is a $q$-dimensional vector of basis functions and $\boldsymbol{\beta}_j$ ($j=1,\ldots, k$) are output coordinate specific coefficients. Similarly, the unknown coordinate-wise variances, $\sigma^2_j$, vary for each $j$. However, the assumption that the range parameters, ${\boldsymbol{\lambda}}$, are shared between different output dimensions leads to a reduction in computational complexity for the emulator.

Parallel partial emulation utilizes the objective prior
\begin{align}
    \pi^R(\boldsymbol{\beta}_1,\dots,\boldsymbol{\beta}_k, \sigma^2_1,\dots,\sigma^2_k) \propto \frac{1}{\prod_{j=1}^k \sigma_j^2}, \label{eq:prior}
\end{align}
for the mean parameters $\boldsymbol{\beta}$ and the variance parameters $(\sigma_1^2,\dots,\sigma^2_k)$. Using common range parameters ${\boldsymbol{\lambda}}$ results in the GP at a new input $\bx^*$ being a multivariate $t$-distribution with dimensional covariance in output components of the form,
\begin{align}
    \mathbf{y}(\bx^*) \mid \bx^\D, Y^\mathcal{D}, {\boldsymbol{\lambda}} \sim t_{n-q}\big{(}\hat{\mathbf{y}}(\bx^*), \mathrm{diag}(\hat{\boldsymbol{\sigma}}^2)c^{**}\big{)}, \label{eq:yxstar_dist}
\end{align}
with
\begin{align}
    \hat{\mathbf{y}}^\top(\bx^*) &= \mathbf{h}(\bx^*) \hat{\boldsymbol{\beta}} + \mathbf{r}^T(\bx^*)\mathbf{R}^{-1}\Big{(}Y^\mathcal{D} - \hxD \hat{\boldsymbol{\beta}}\Big{)}, \label{eq:yhatxstar_mean_PPE}
\end{align}
and $\hat{\boldsymbol{\sigma}}^2 = (\hat{\sigma}_1^2,\dots,\hat{\sigma}_k^2)$. Here $\hat{\sigma}_j^2$ is defined as, 
\begin{align}
    \hat{\sigma}_j^2 &= (n-q)^{-1}\big{(}\mathbf{y}^\mathcal{D}_j - \hxD\hat{\boldsymbol{\beta}_j}\big{)}^\top \mathbf{R}^{-1}\big{(}\mathbf{y}^\mathcal{D}_j - \hxD\hat{\boldsymbol{\beta}}_j\big{)}, \label{eq:sigmahat2_PPE}
\end{align}
and $c^{**}$ is defined the same as in Eq. \ref{eq:c**} and $\mathbf{y}^\mathcal{D}_j$ is the $j^{th}$ column of $Y^\mathcal{D}$. $\hat{\boldsymbol{\beta}}$ is a matrix of mean trend parameter estimates, where column $\hat{\boldsymbol{\beta}}_j$ is the mean trend parameter estimate for output $j$ and is defined as, 
\[
\hat{\boldsymbol{\beta}_j} = \big{(}\htxD\mathbf{R}^{-1} \hxD\big{)}^{-1} \htxD\mathbf{R}^{-1}\mathbf{y}^\mathcal{D}_j,
\] 
the generalized least squares estimate for $\boldsymbol{\beta}_j$.


Gu and Berger showed that the predictive mean of the parallel partial GP emulator at a new input $\bx^*$, $\hat{\mathbf{y}}(\bx^*)$ can be expressed as a weighted sum of the simulator outputs $\hat{\mathbf{y}}(\bx^*) = \boldsymbol{\omega}(\bx^*)Y^\mathcal{D}$ with,
\begin{align}
\boldsymbol{\omega}(\bx^*)=\big{(}\hxstar - \mathbf{r}^\top(\bx^*)\mathbf{R}^{-1}\hxD\big{)}\big{(}\htxD\mathbf{R}^{-1}\hxD\big{)}^{-1} \htxD\mathbf{R}^{-1} + \mathbf{r}^\top(\bx^*)\mathbf{R}^{-1}. \label{eq:weights}
\end{align}
This result leads to the parallel partial emulator not only interpolating the design inputs but the PPE is also a weighted sum of the simulator output over all $k$ dimensions. The full computational cost of fitting the emulator is $\mathcal{O}(tn^3) + \mathcal{O}(tn^2 k)$ \citep{Gu:Berg:2016}. Here, $t$ is the number of iterations (posterior evaluations in an optimization routine) in the range parameter estimation process, $n$ is the number of model runs, and $k$ is the number of output dimensions. Thus the computational cost of the PPE is linear in $k$, leading to the reduction in computational complexity. Note, the \texttt{RobustGaSP} package can take scalar or vector valued outputs for fitting GPs or PPEs. The computational cost of $\mathcal{O}(tn^3) + \mathcal{O}(tn^2 k)$ can still be very expensive when the number of model runs, $n$, is large. In the following section look to utilize Scaled Vecchia Approximation within the PPE framework to make this approach more tractable.

\section{The Veccia Parallel Partial Emulator\label{sec:Methods}}
Here we derive the new {\it Vecchia parallel partial emulator} (VPPE).  The VPPE method allows for GP models of (potentially very large) multidimensional output data that are computationally feasible for a large number of model runs. To facilitate fitting the VPPE, we begin by deriving the new {\it Vecchia marginal posterior} that combines the Scaled Vecchia approximation \citep{Katz:Guin:Lawr:2022} with the Robust GaSP approach to fitting GP range parameters \citep{Gu:Wang:Berg:2018}. We also confirm that the Vecchia marginal posterior is the marginal posterior when the ``nearest neighbors" are all the neighbors (i.e., when $m=n-1$). We further derive the gradient of new the Vecchia marginal posterior for use within optimization schemes for estimating the range parameters with mode posteriors.  We refer to fitting scalar GPs with this approach as {\it Scaled Vecchia Robust GaSP} (VRGaSP) and when applied to multi-dimensional output, VPPE. Finally, we explore the computational complexity of the VPPE.

\subsection{The Vecchia marginal posterior \label{sec:int_lik_section}}
The Robust GaSP approach to fitting GPs relies on mode posterior estimates of range parameters, ${\boldsymbol{\lambda}}$, where the posterior contains a likelihood that is marginalized over $\boldsymbol{\beta}$ and $\sigma^2$ in order to account for uncertainty in those parameters \citep{Gu:Wang:Berg:2018}. Alternatively, the Scaled Vecchia Approximation, \citep{Katz:Guin:Lawr:2022}, maximizes the profiled likelihood, $L^*_m(\blambda \mid \bx^\D, \by^\D, \bbeta_m, \sigma^2_m)$ with generalized least squares estimates of   the mean trend parameters and the scalar variance, $\bbeta_m$ and $\sigma^2_m$, respectively \citep{Vecchia:1988}.

To derive the Vecchia marignal posterior, we begin by multiplying the Vecchia likelihood, $L_m(\blambda, \boldsymbol{\beta}, \sigma^2 \mid \bx^\D, \by^\D)$, given in Eq.~\ref{eq:Vecchia_lik}, by the standard reference prior for $\bbeta$ and $\sigma^2$, $\pi^R(\boldsymbol{\beta}, \sigma^2) \propto \frac{1}{\sigma^2}$. We then integrate this product over $\boldsymbol{\beta}$ and $\sigma^2$. This Vecchia marginal likelihood is thus given by,
\begin{align}
\mathcal{L}_m(\boldsymbol{\lambda} \mid \bx^\D, \by^\D) &= \int_{\mathbb{R}^q \times (0,\infty)} L_m(\blambda, \boldsymbol{\beta}, \sigma^2 \mid \bx^\D, \by^\D) \cdot \frac{1}{\sigma^2} d\boldsymbol{\beta} d\sigma^2, \nonumber\\
   &\propto \int_{\mathbb{R}^q \times (0,\infty)} (\sigma^2)^{-n/2} \left( \prod_{i=1}^n \omega_{im}^{-1/2} \right) \exp \left\{-\frac{1}{2\sigma^2} \sum_{i=1}^n \omega_{im}^{-1}e_{im}^2 \right\} \cdot \frac{1}{\sigma^2} d\boldsymbol{\beta} d\sigma^2, \label{eq:vecchia_beta_integral}\\
   &\propto \left( \prod_{i=1}^n \omega_{im}^{-1/2} \right) |\tilde{\mathrm{\Sigma}}|^{-1/2} \int_0^\infty (\sigma^2)^{-\frac{(n-q)}{2}-1}  \exp \left\{-\frac{\tilde{S}^2}{2\sigma^2}\right\} d\sigma^2,\label{eq:vecchia_sigma2_integral}\\   
   &\propto \left( \prod_{i=1}^n \omega_{im}^{-1/2} \right) |\tilde{\mathrm{\Sigma}}|^{-1/2} \left(\tilde{S}^2\right)^{-\frac{n-q}{2}},\label{eq:result}
\end{align}
where
\begin{align}
    \tilde{\mathrm{\Sigma}} &= \sum_{i=1}^n\tilde{\mathbf{h}}_{im}^\top\omega_{im}^{-1}\tilde{\mathbf{h}}_{im},\label{eq:Sigma_tilde}\\
    \tilde{S}^2 &= \sum_{i=1}^n g_{im} \omega_{im}^{-1} g_{im} - \left[ \sum_{i=1}^n \tilde{\mathbf{h}}_{im}^\top\omega_{im}^{-1}g_{im} \right]^\top \left[\sum_{i=1}^n\tilde{\mathbf{h}}_{im}^\top\omega_{im}^{-1}\tilde{\mathbf{h}}_{im} \right]^{-1}\left[ \sum_{i=1}^n \tilde{\mathbf{h}}_{im}^\top\omega_{im}^{-1}g_{im} \right], \label{eq:Ssqrd}\\
    g_{im} &= y_i - \mathbf{r}_{im}^\top(\bR_{im}+\nu^2I)^{-1}\mathbf{y}_{im}, \nonumber \\
    \tilde{\mathbf{h}}_{im} &= \mathbf{r}_{im}^\top(\bR_{im}+\nu^2I)^{-1}\mathbf{h}(x_{im}) - \mathbf{h}(x_i), \nonumber
\end{align}
and all other terms are as they were defined in Eqs.~\ref{eq:vec_cond_var}-\ref{eq:vec_corr_im}. Details of this derivation are shown in the Appendix (\ref{sec:appendix}). 

Finally, we multiply the Vecchia marginal likelihood by the reference prior for the range parameters, $\pi(\blambda)$ ($\pi^R(\boldsymbol{\lambda})$ or $\pi^{JR}(\boldsymbol{\lambda})$) to obtain the Vecchia marginal  posterior,
\begin{align}
    \pi_m(\boldsymbol{\lambda} \mid \bx^\D, \mathbf{y}^\D) &\propto \left( \prod_{i=1}^n \omega_{im}^{-1/2} \right) |\tilde{\mathrm{\Sigma}}|^{-1/2} \left(\tilde{S}^2\right)^{-\frac{n-q}{2}}\pi(\boldsymbol{\lambda}) . \label{eq:post_result}
\end{align}
Recall the marginal likelihood from Eq.~\ref{eq:likelihood}, 
\begin{align*}
\mathcal{L}(\boldsymbol{\lambda} \mid \bx^\D, \mathbf{y}^\mathcal{D}) \propto |\mathbf{R}|^{-1/2}|\htxD \mathbf{R}^{-1} \hxD|^{-1/2} \left(S^2 \right)^{-\left(\frac{n-q}{2}\right)}.
\end{align*}
When $m=n-1$, if we compare the terms in the Vecchia marginal likelihood to those in $\mathcal{L}(\boldsymbol{\lambda} \mid \bx^\D, \mathbf{y}^\mathcal{D})$, we see that
\begin{align*}
    \prod_{i=1}^n \omega_{im}^{-1/2} &= | \mathbf{R} |^{-1/2},\\
    |\tilde{\mathrm{\Sigma}}|^{-1/2} &= |\htxD \mathbf{R}^{-1} \hxD|^{-1/2},\\
    \tilde{S}^2 &= S^2.
\end{align*}
Thus, when $m=n-1$, $\mathcal{L}_m(\boldsymbol{\lambda} \mid \bx^\D, \mathbf{y}^\mathcal{D})=\mathcal{L}(\boldsymbol{\lambda} \mid \bx^\D, \mathbf{y}^\mathcal{D})$, and hence $\pi_m(\boldsymbol{\lambda} \mid \bx^\D, \mathbf{y}^\D)=\pi(\boldsymbol{\lambda} \mid \bx^\D, \mathbf{y}^\D)  $.
This proof is also shown in the Appendix (\ref{sec:appendix_equivalence}).

\subsection{Derivation of the gradient of the Vecchia marginal posterior}
Many optimization schemes, including  L-BFGS \citep{Nocedal:1980, Liu:Nocedal:1989, Ypma:2014} which we utilize in numerical experiments in Section~\ref{sec:Results}, require gradient computations of the  loglikelihood. Here we derive the gradient for the Vecchia marginal loglikelihood. The profiled loglikelihood in \cite{Guiness:2021} is given as,
\begin{align}
    -2\log L^*_m \propto n \sigma_m^2 + \sum_{i=1}^n \log \left(\omega_{im}\right), \label{eq:vecchia_log_lik_profiled}
\end{align}
with
\begin{align*}
    \sigma_m^2 &= \frac{1}{n}\sum_{i=1}^n \omega_{im}^{-1} \tilde{e}_{im}^2,\\
    \boldsymbol{\beta}_m &= \left[\sum_{i=1}^n \omega_{im}^{-1}\tilde{\mathbf{h}}_{im} \tilde{\mathbf{h}}_{im}^\top \right]^{-1} \left[ \sum_{i=1}^n \omega_{im}^{-1}\tilde{\mathbf{h}}_{im} g_{im}\right],\\
    \tilde{\mathbf{h}}_{im} &= \mathbf{r}_{im}^\top(\bR_{im}+\nu^2I)^{-1}\mathbf{h}(x_{im}) - \mathbf{h}(x_i),\\
    g_{im} &= y_i - \mathbf{r}^\top_{im}(\bR_{im}+\nu^2 I)^{-1}\mathbf{y}_{im},
\end{align*}
and $\tilde{e}_{im}$ is identical to $e_{im}$ in Eq. \ref{eq:vec_cond_residual} except replacing $\boldsymbol{\beta}$ with $\boldsymbol{\beta}_m$. The $j^{th}$ component of the gradient of this Vecchia profiled loglikelihood is,
\begin{align}
    \frac{\partial}{\partial \lambda_j} \left(-2\log {L}^*_m \right) \propto \frac{\partial}{\partial \lambda_j} \left(n \sigma_m^2 \right) + \frac{\partial}{\partial \lambda_j} \left(\sum_{i=1}^n \log \left(\omega_{im}\right) \right). \label{eq:vecchia_log_lik_profiled_gradient}
\end{align}
Our Vecchia marginal loglikelihood is defined slightly differently (note that $n\sigma^2_m = \tilde{S}^2$ as shown in Eq.~\ref{eq:sigmisS}),
\begin{align}
     -2\log \mathcal{L}_m &\propto (n-q) \log\left(\tilde{S}^2\right) + \sum_{i=1}^n \log \left(\omega_{im}\right) + \log \left|\tilde{\mathrm{\Sigma}} \right|, \nonumber\\
     &= (n-q) \log\left(n\sigma_m^2\right) + \sum_{i=1}^n \log \left(\omega_{im}\right) + \log \left|\sum_{i=1}^n \tilde{\mathbf{h}}_{im}^\top \omega_{im}^{-1}\tilde{\mathbf{h}}_{im} \right|.\label{eq:vecchia_log_lik_int}
\end{align}
The subsequent derivative of this loglikelihood with respect to $\lambda_j$ is
\begin{align}
     \frac{\partial}{\partial \lambda_j} \left(-2\log \mathcal{L}_m \right) \propto (n-q) \frac{\frac{\partial}{\partial \lambda_j} \left(n\sigma_m^2\right)}{n\sigma_m^2} + \frac{\partial}{\partial \lambda_j} \left(\sum_{i=1}^n \log \left(\omega_{im}\right) \right) + \frac{\partial}{\partial \lambda_j} \left(\log \left|\sum_{i=1}^n \tilde{\mathbf{h}}_{im}^\top \omega_{im}^{-1}\tilde{\mathbf{h}}_{im} \right| \right).\label{eq:vecchia_log_lik_int_gradient}
\end{align}
We can see in Eq.~\ref{eq:vecchia_log_lik_int_gradient} that there are two changes to the gradient of the profile likelihood in Eq.~\ref{eq:vecchia_log_lik_profiled_gradient}. The first is that the derivative of $n\sigma_m^2$ becomes the ratio of the derivative of $n\sigma^2_m$ over $n\sigma^2_m$. The second difference is  the derivative of the additional term, $\log |\sum_{i=1}^n \tilde{\mathbf{h}}_{im}^\top \omega_{im}^{-1}\tilde{\mathbf{h}}_{im} |$. (Note, from Eq.~\ref{eq:Sigma_tilde},  $\log | \tilde{\mathrm{\Sigma}} | = \log |\sum_{i=1}^n \tilde{\mathbf{h}}_{im}^\top \omega_{im}^{-1}\tilde{\mathbf{h}}_{im} |$.) For $j=1,\dots,p$, the partial derivative of $\log | \tilde{\mathrm{\Sigma}} |$ with respect to $\lambda_j$ simplifies to,
\begin{align*}
    \frac{\partial}{\partial \lambda_j} \log| \tilde{\mathrm{\Sigma}} | &= \frac{\frac{\partial}{\partial \lambda_j} |\tilde{\mathrm{\Sigma}} |}{| \tilde{\mathrm{\Sigma}} |},\\
    &= \frac{|\tilde{\mathrm{\Sigma}}| \cdot \mathrm{tr}\left(\tilde{\mathrm{\Sigma}}^{-1} \cdot \frac{\partial}{\partial \lambda_j} \tilde{\mathrm{\Sigma}}\right)}{|\tilde{\mathrm{\Sigma}}|},\\
    &= \mathrm{tr}\left(\tilde{\mathrm{\Sigma}}^{-1} \cdot \frac{\partial}{\partial \lambda_j} \tilde{\mathrm{\Sigma}}\right).
\end{align*}
Here $\frac{\partial}{\partial \lambda_j} \tilde{\mathrm{\Sigma}}$ is defined as the elementwise derivatives of $\tilde{\mathrm{\Sigma}}$ with respect to $\lambda_j$. Further details on this computation can be found in \cite{Guiness:2021}.

\subsection{The Vecchia marginal likelihood for multidimensional output}
PPE is well suited to problems that have large output dimensions, whereas Vecchia approximations are well suited to problems with a large number of model runs. Here, we combine the Scaled Vecchia approximation method \citep{Katz:Guin:Lawr:2022} with the PPE~\citep{Gu:Berg:2016} to develop the VPPE method. 

In parallel partial emulation, independent GPs (with a shared correlation structure over input space) are developed for each of a simulator's output dimensions  (e.g.~Eq.~\ref{eq:yxstar_dist}). Thus, the extension of the Vecchia marginal likelihood to multidimensional outputs is straightforward, as we will simply compute a separate Vecchia marginal likelihood for each output dimension. The definitions for terms $\tilde{\mathrm{\Sigma}}$ and $\tilde{{S}}^2$ are shown in Section \ref{sec:int_lik_section}. We observe that $\tilde{S}^2$ from Eq. \ref{eq:Ssqrd} is the only expression that contains output values. Instead of computing a scalar value for $\tilde{S}^2$, we compute a vector of dimension $k$, one for each simulator output dimension, where each dimension corresponds to the scalar $\tilde{S}^2$ value for the output at that dimension.

Similar to the PPE approach, we compute the marginal likelihood for multidimensional outputs as the product of the individual marginal likelihoods for each dimension, but we do so with Vecchia marginal likelihoods. That is, for a $k$-dimensional output, and resulting vector $\mathbf{\tilde{S}^2}$, the likelihood will be as follows,
\begin{align}
    \mathcal{L}_m(\blambda \mid \bx^\D, Y^D) \propto \left( \left( \prod_{i=1}^n \omega_{im}^{-1/2} \right) \left| \tilde{\mathrm{\Sigma}} \right|^{-1/2} \right)^k \prod_{l=1}^k \left(\tilde{S}^2_l\right)^{-\frac{n-q}{2}},
\end{align}
where $\tilde{S}^2_l$ is the $l^{th}$ entry of vector $\mathbf{\tilde{S}^2}$.

\subsection{Computational complexity}
PPE has a computational complexity of $\mathcal{O}(n^3) + \mathcal{O}(n^2k)$ \citep{Gu:Berg:2016}. While finding the estimates for the range parameters ${\boldsymbol{\lambda}}$, two computationally intensive steps occur during each iteration of the optimization process: the computation of a new ${\mathbf{R}}^{-1}$, which has a computational cost of $\mathcal{O}(n^3)$, and the evaluation of the likelihood in Eq. \ref{eq:likelihood}, which has a computational cost of $\mathcal{O}(n^2k)$. Since VPPE utilizes a composition of $n$ separate $m\times m$ matrices to approximate the full $\mathbf{R}$, VPPE has a computational complexity of $\mathcal{O}(nm^3) + \mathcal{O}(m^2k)$. 

With the Vecchia marginal posterior extended to multidimensional outputs, we are now able to combine the flexibility and robust nature of PPE to the computational efficiency of the Vecchia approximation when the number of model runs is large. 

\section{Numerical Experiments and Results}\label{sec:Results}
In this section we explore applying VRGaSP and VPPE to three increasingly realistic computer models. We analyze the performance of parallel partial emulation when using the Scaled Vecchia approximation of the likelihood function within the PPE posterior. Throughout the section we will use ``PPE (RGaSP)" to refer to the original method with the full marignal likelihood function evaluation for parallel partial emulation (scalar GP emulation) and ``VPPE (VRGaSP)" to refer to the method where we use the Vecchia marginal likelihood function within the parallel partial emulator (scalar GP emulator) posterior.
While fitting the PPE to the large-$n$ training data sets often takes over an hour, fitting the VPPE to the same data sets takes only a few minutes or less. Yet in simulation experiments, the predictive RMSE for VPPE remains comparable to PPE even as $n$ increases demonstrating VPPE's drastic improvement in computational efficiency without having to sacrifice accuracy.

The first ``simulator" data set consists of samples from a prescribed Gaussian process. Here our experiments involve fitting emulators with both PPE/RGaSP and VPPE/VRGaSP methods. Further we explore their computational efficiency and their predictive skill. In these experiments, we will vary both the number of training model runs $n$, the number of nearest neighbors $m$, and the number of output dimensions $k$.

The second simulator, known as Richard's Equation, is a partial differential equation (pde) model of water flow through a soil column.  In this hydrology simulator, we will show the accuracy of emulators fit using VPPE method as compared to PPE. Further, we will show the efficacy of VPPE over the perhaps more common usage of Vecchia that involves reformatting multidimensional spatial output data to scalar-output data with column depth treated as an input dimension.

Finally, we consider a volcanic flow simulator with a very large number of spatial output dimensions -- e.g.~spatially varying inundation depth for a given flow scenario (volume, vent opening location, and mobility parameters.) Such flows are highly topologically driven, and as such a large number of potential vent locations must be explored in a simulation study. Hence we have a large, spatially dependent training set and a very large number of output dimensions. Again, VPPE is much faster than PPE, both for fitting and prediction while yielding equivalent predictions.

\subsection{Synthetic computer model example}
The first ``simulator" example we explore is a simulated multivariate Gaussian Process. We fix a set of range parameters, $\boldsymbol{\lambda}$, and choose the Matérn 3/2 covariance function for simulating our output data. We take our output data to be realizations of a GP,
\begin{align}
    y(\cdot) &\sim \mathcal{GP}(\mu(\cdot), C(\cdot,\cdot)),\\
    \mathbf{R}_{ij} &= \prod_{l=1}^p c_{3/2}(\bx_i^\mathcal{D},\bx_j^\mathcal{D}; \lambda_l),
\end{align}
with a specified variance $\sigma^2$ and $c_{3/2}(\cdot,\cdot)$ referring to the Matérn 3/2 correlation function. We simulate these GP ``model runs" for both the scalar output ($k=1$) and vector output ($k=100$) cases.

To simulate the GPs, we first perform random Latin Hypercube sampling over a $p$-dimensional unit cube to obtain our design, $\bx^\mathcal{D}$, from the package \verb|lhs| \citep{Carnell:2024} with $2n$ design points and $p$ input dimensions. We then calculate the covariance matrix $\sigma^2 \mathbf{R}$ of these design points with a set of specified range parameters. Our simulated output data is obtained by sampling $\mathcal{N}(0,1)$ random variables to fill in a $2n \times k$ matrix $U$ and transforming it via, $\mathbf{y} = L^\top U$,
where $L$ is the Cholesky factor of the covariance matrix $\sigma^2 \mathbf{R}$.

Once we have generated $2n$ model runs, we split them into $n$ training runs and $n$ testing runs. For both experiments and for both $k=1$ and $k=100$, we fix the number of input dimensions $p$ at 4 with range parameters $(0.5,0.8,1.2,0.3)$. We also choose $\sigma^2=1$. For the first experiment, we set the number of nearest neighbors, $m$, in VPPE to be fixed at 30 and increase $n$ from $n=100$ to $n=4000$. For the second experiment, we fix $n=4000$ and vary $m$ in VPPE from $m=5$ to $m=100$. For both experiments we recorded the times taken to fit the emulator, along with the RMSE on the output data. All of the timing results were obtained on a desktop computer (3.6 GHz Intel 8 Cores i7-9700K).

\subsubsection{Computer experiment 1: fix $m$, vary $n$.}
\begin{figure}[htbp]
    \centering
    \includegraphics[width=0.99\linewidth]{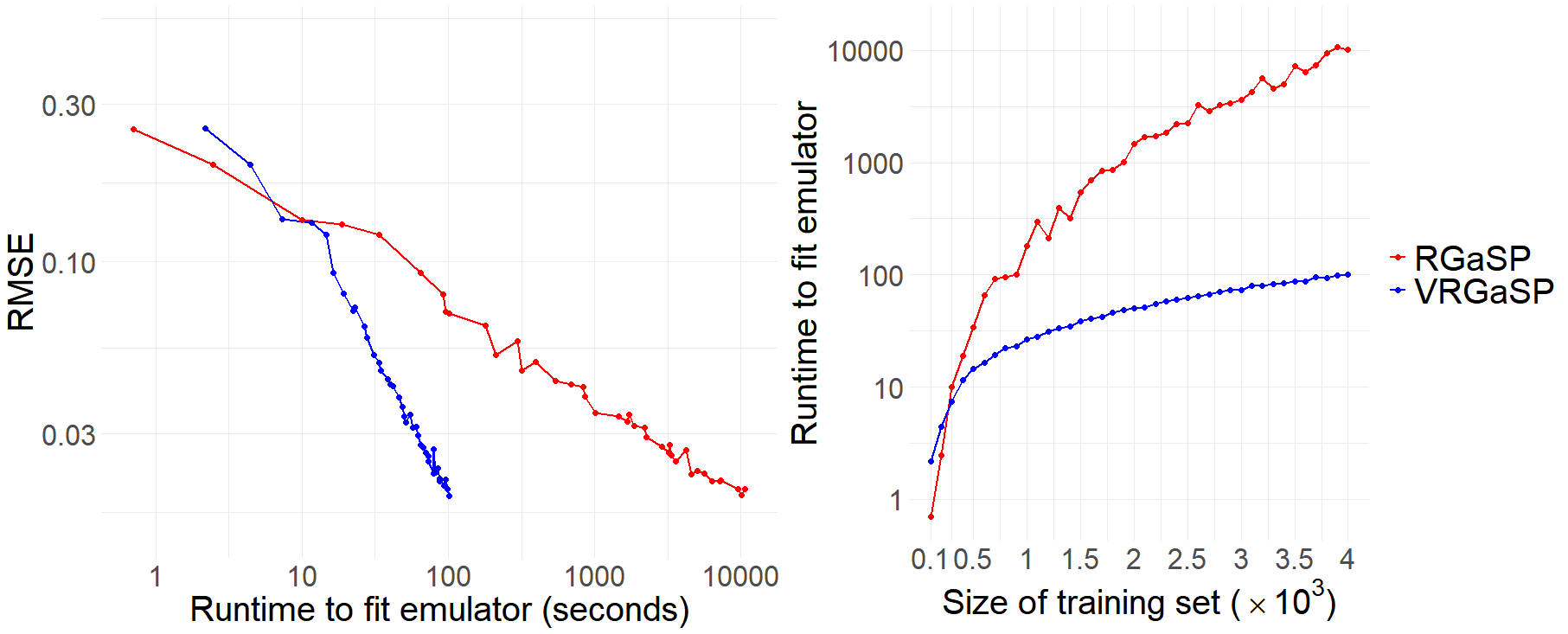}
    \caption{Left: relative RMSE vs.~runtime to fit the emulator for varying size of training sets ($100 \leq n \leq 4000$), with 4 input dimensions and 1 output dimension. Right: runtime vs.~size of training sets. For VRGaSP, $m$ was fixed at 30 and had a maximum runtime of 101 seconds. The maximum runtime for RGaSP was 10,600 seconds (when $n = 3900$). The minimum relative RMSE for both methods was 0.0195.}
    \label{fig:GP_runtimes_varying_n_k_1}
\end{figure}

\paragraph{Scalar output ($k=1$):} First we will fit Scaled Vecchia RobustGaSP (VRGaSP) and Robust GaSP (RGaSP) to simulated data with a single output dimension $k=1$. As discussed in Section \ref{sec:Methods}, VRGaSP has a significantly lower computational cost when $m \ll n$. We see this reflected in the runtimes in Figure~\ref{fig:GP_runtimes_varying_n_k_1}.
Figure~\ref{fig:GP_runtimes_varying_n_k_1} shows the relative root mean squared error (RMSE) vs the runtime to fit the emulator for both RGaSP and VRGaSP as the size of the training set increases from $n=100$ to $n=4000$. Here, the RMSE is the ratio of the root mean squared error of each emulator's predicted output to the simulated ``true" output.
While VRGaSP's runtime increases linearly with $n$, RGaSP increases cubically, leading to very long runtimes quite quickly as $n$ increases. It is important to note that for small $n$, the difference in computational complexity for RGaSP and VRGaSP is practically negligible. For this reason, the VRGaSP's computational advantage is realized for $n$ greater than approximately $300$.

Additionally, we notice signs of instability in the L-BFGS optimization routine as $n$ surpasses roughly 500. Fitting a GP in the \texttt{RobustGaSP} package requires two runs the optimization routine, one seeded with ``small" initial values for range parameters and one with ``large" initial values. Despite employing the same twice-seeded L-BFGS optimization routine, VRGaSP had nearly no issues with L-BFGS convergence regardless the number of training runs. On the other hand, L-BFGS within RGaSP becomes more dependent on the initialization values as $n$ increases, leading to inconsistent convergence. That is, L-BFGS sometimes fails to converge and instead outputs the seed values for range parameter estimates. When $n$ surpasses roughly 500, the first optimization with larger initial range parameters often fails to converge within RGaSP, but will consistently converge for both optimizations within VRGaSP. Typically when the RGaSP optimization does not converge, the optimization process is cut short, leading to faster runtimes. This convergence issue causes the RGaSP curves of runtime vs size of training set to appear somewhat jagged, while VRGaSP remains smooth throughout the experiments, as we can see in Figure~\ref{fig:GP_runtimes_varying_n_k_1}. We believe that the large-$n$ L-BFGS convergence issue is likely due to numerical instability that starts to appear when computing Cholesky decompositions of large matrices which are used within the L-BFGS optimization routine \citep{Scha:Katz:Owha:2021, Marc:Zing:2023}. While the full correlation matrix $\mathbf{R}$ (Section \ref{sec:GP}) has size $n \times n$, Vecchia's matrices $\bR_{im}$ (\ref{eq:vec_corr_im}) are of size $m \times m$, which are typically defined to be small, ensuring that this numerical instability issue is often avoided.

\begin{figure}[htbp]
    \centering
    \includegraphics[width=0.99\linewidth]{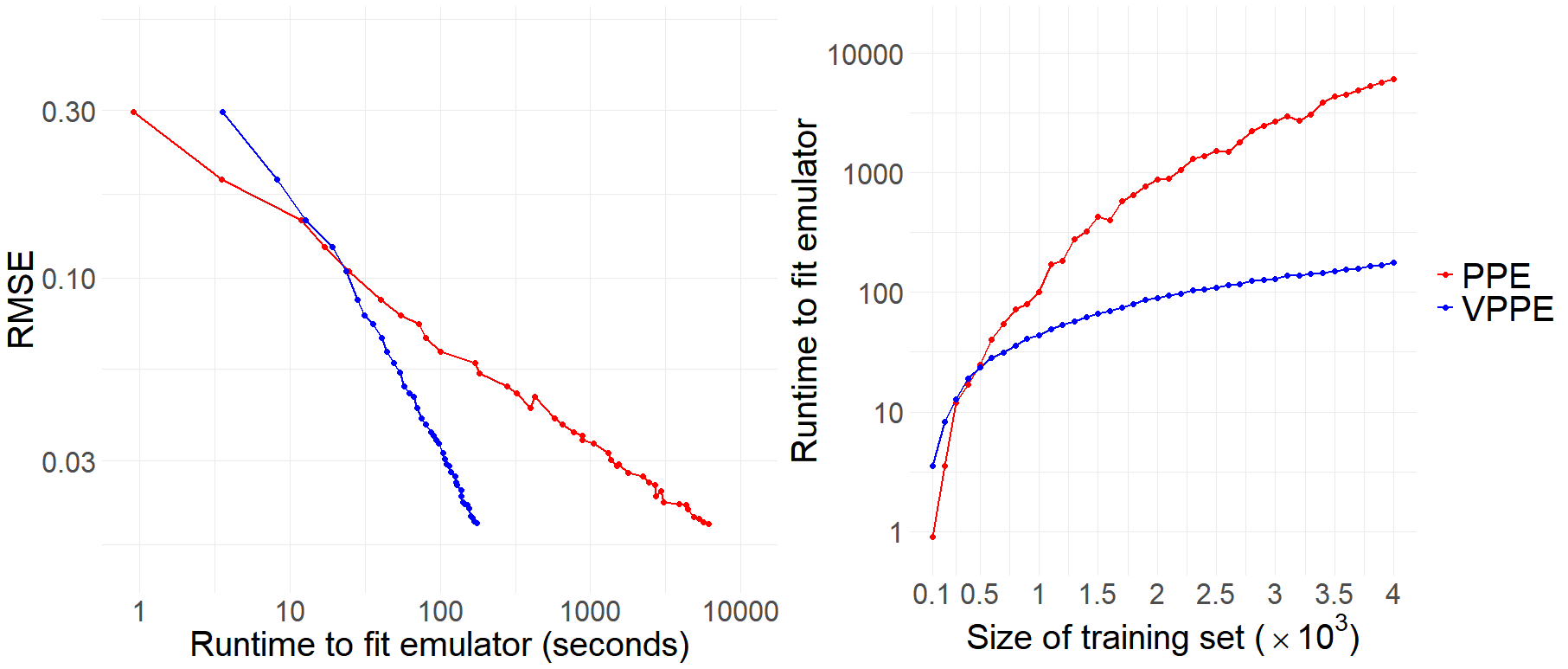}
    \caption{Left: relative RMSE vs.~runtime to fit the emulator (left) for varying size of training sets ($100 \leq n \leq 4000$), with 4 input dimensions and 100 output dimensions. Right: runtime vs.~size of training sets. For VPPE, $m$ was fixed at 30. The maximum runtime for VPPE was 176 seconds and the maximum runtime for PPE was 6100 seconds. The minimum relative RMSE for both methods was 0.0199.}
    \label{fig:GP_runtimes_varying_n}
\end{figure}

\paragraph{Vector output ($k=100$):} Now we perform the same experiment, but for $k=100$. As anticipated, we see that the VPPE has a significantly lower computational cost than PPE when $m \ll n$.  Figure~\ref{fig:GP_runtimes_varying_n} shows similar results to the scalar-output case where the efficiency of fitting the VPPE over PPE becomes dramatic for $n>500$ with comparable predictive skill for a given $n$. For $n=4000$, PPE took well over an hour estimating the optimal range parameters, while VPPE took less than three minutes (176 seconds).
 This experiment demonstrates that the VPPE's approximation to fitting the PPE with the Vecchia marginal likelihood function is successful, leading to similar RMSEs when compared to PPE.

Note that for most $n$, RGaSP took {\it longer} in the $k=1$ case when compared to PPE in the $k=100$ case. This unintuitive discrepancy is most likely due to fact that the optimization process proceeded to convergence more often in the $k=1$ case than in the $k=100$ case. That is, in the $k=100$ case, L-BFGS within PPE would often not converge leading to the optimization process getting cut short. This did not happen for $k=1$ nearly as often, so the optimization process would take longer overall than for $k=100$ even though the iterations within L-BFGS were slightly faster for the $k=1$ case. On the other hand, VRGaSP and VPPE's optimization processes converged consistently for $k=1$ and $k=100$, utilizing roughly the same number of iterations per optimization in each case. Thus the VRGaSP is faster than VPPE for the same $n$, as one would anticipate.

\subsubsection{Computer experiment 2: fix $n$, vary $m$.}
\begin{figure}[htbp]
    \centering
    \includegraphics[width=0.99\linewidth]{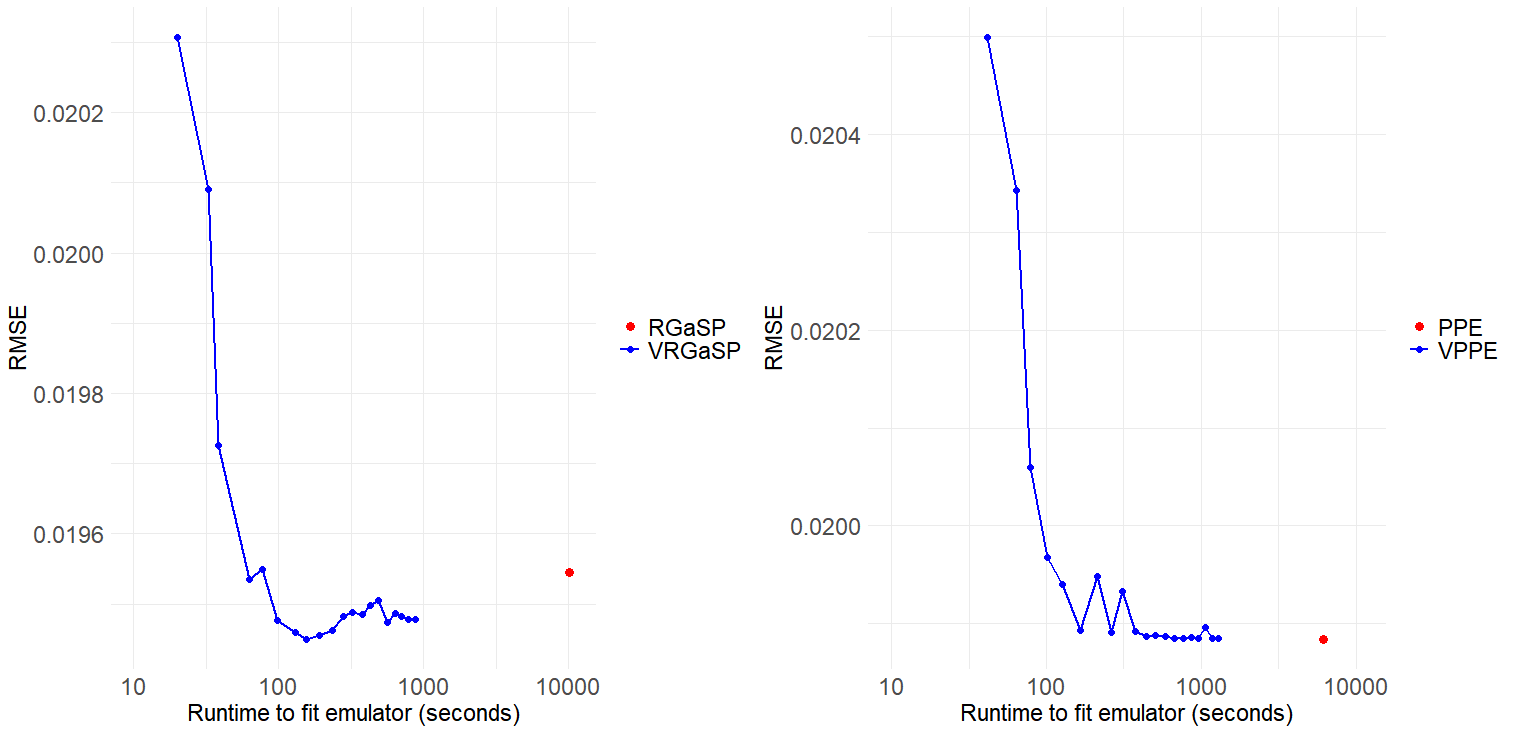}
    \caption{RMSE vs.~runtime (on a log scale) to fit the emulator for varying size of nearest neighbors ($5 \leq m \leq 100$) in VRGaSP/VPPE. For both RGaSP/VRGaSP (left) and PPE/VPPE (right), we fixed $n=4000$ data points in the training set.}
    \label{fig:GP_RMSE_v_runtimes_m}
\end{figure}

While we have observed that choosing $m=30$ leads to consistent and accurate GP fitting based on predictive performance, $m$ can vary as the user sees fit. Here we again consider $p=4$ input dimensions and set the training size to $n=4000$. We vary the nearest neighbors $m$ by 5 from $m=5$ to $m=100$ and calculate predictive relative RMSEs for RGaSP/VRGaSP and PPE/VPPE. 
\paragraph{Scalar output ($k=1$):}
 In Fig. \ref{fig:GP_RMSE_v_runtimes_m} we plot the predictive RMSE vs.~runtime, and we can see that choosing $m$ as low as 5 can still yield fairly low predictive RMSEs. In fact, we noticed that VRGaSP consistently outperforming RGaSP for $m \geq 30$, despite taking less than 1\% of the runtime for $m=30$ and at less than 9\% for $m=100$. The time discrepancy between VRGaSP and RGaSP is quite large in this scalar-output case. The maximum runtime for VRGaSP was 876 seconds and the runtime for RGaSP was 10,100 seconds. The RMSE for RGaSP was 0.0195, and the minimum RMSE for VRGaSP was 0.0194 when $m= 40$ (with a runtime of 156 seconds). 

\paragraph{Vector output ($k=100$):}
When performing this same experiment for $k=100$, we notice similar results to the scalar-output case. While VPPE does not outperform PPE for lower $m$ like in the scalar-output case, the predictive accuracy becomes comparable for $m \geq 30$, despite taking only a fraction of the runtime. The maximum runtime for VPPE was 1300 seconds and the runtime for PPE was 6100 seconds. The RMSE for PPE was the same as VPPE for $m\geq 30$ and was 0.0199.



The range of the testing outputs was from -4.20 to 4.38, leading to the RMSE accounting for about 0.2\% of the range of the testing output values. We can see that the VPPE approximation leads to the emulator producing near identical results to PPE even with the VPPE has very few nearest neighbors. In fact, using as little as 30 neighbors for VPPE leads to the same RMSE as PPE within 4 decimal places despite taking less than 3\% of the runtime.

\subsection{Hydrology model}
The Richards' equation \citep{Richards:1931} models subsurface water flow in hydrology. We will consider a version of Richard's equation that incorporates the Brooks-Corey model \citep{Salv:Ente:1994}. 
\begin{align}
    \frac{\partial \theta}{\partial t} = \nabla \cdot \left( K(\theta) \frac{\partial \psi}{\partial \theta} \nabla \theta \right), \label{eq:richard}
\end{align}
with
\begin{align}
    K(\theta)=K_s \left(\frac{\psi}{\psi_b}\right)^{-2-3\lambda}, \qquad \psi = \psi_b \left(s_e^{-\frac{1}{\lambda}} \right), \qquad s_e = \frac{\theta-\theta_r}{\theta_s-\theta_r}. \label{eq:BCmodel}
\end{align}
In this simulator, we consider three input parameters: $\theta_s$, the porosity, $K_s$, the saturated hydraulic conductivity, and $\lambda$, the pore-size distribution parameter. The output is $\theta$, the volumetric water content as a function of (1-D) soil depth. The fixed parameters in the equation are $\psi$, the pressure head of soil, $K(\theta)$, the hydraulic conductivity, $\psi_b$, the air-entry pressure, $s_e$, the effective saturation, and $\theta_r$, the residual water content. Note that all simulations considered here are forced with the same rainfall time series. The upper boundary condition is constant flux when forced by rain, and a constant pressure head otherwise. The bottom boundary condition is a context flux that models a the outflow of water from the column.
\begin{figure}[htbp]
    \centering
    \includegraphics[width=0.75\linewidth]{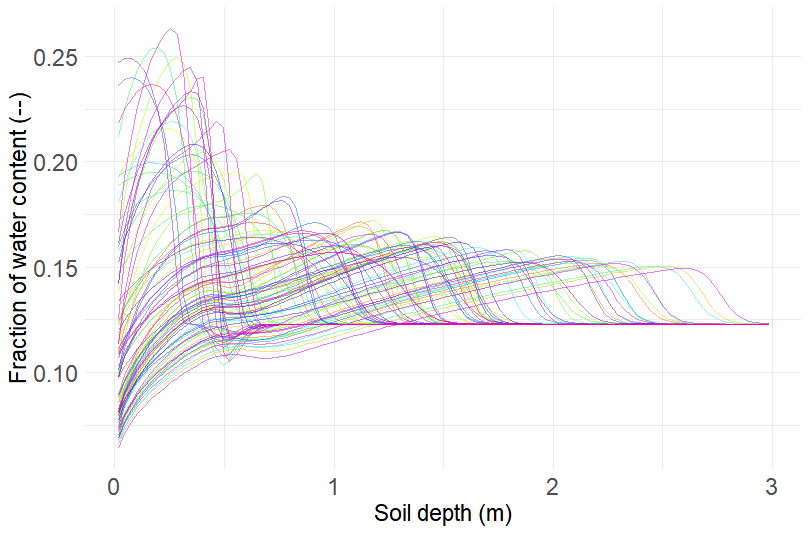}
    \caption{100 output runs of the 1996 simulated solutions to the modified Richards' equation simulator.}
    \label{fig:RICHARD}
\end{figure}

The dataset has $n=1996$ model runs from LHS over $p=3$ input dimensions and $k=100$ output dimensions. The three input dimensions are porosity ($\theta_s$), ranging from 0.3 to 0.5 (--), saturated hydraulic conductivity ($K_s$), ranging from 0.00057 to 0.028 (mm/s), and pore-size distribution parameter ($\lambda$), ranging from 0.32 to 5 (--). The output data is volumetric water content ($\theta$) along the length of the soil column. We use numerical simulations to approximate the solution to Richards' equation over a grid of 100 soil depths ranging from 0.015 meters to 2.985 meters. Some of the simulator water content output curves are plotted in Figure \ref{fig:RICHARD}. Note the strong spatial non-stationary behavior of these solutions. Utilization of VPPE will allow for the construction of an emulator that is comparable to PPE in terms of accuracy at the fraction of a runtime.

To show the accuracy of VPPE when compared to PPE for this simulator, we ran 20 simulation experiments, where we randomly split the dataset into 300 training runs, and 1696 testing runs. We then fit an emulator for both PPE and VPPE in each simulation to observe the accuracy of VPPE compared to PPE. For all 20 experiments, we set the number of nearest neighbors for VPPE to $m=30$. The variation in RMSE came almost entirely from the emulators built on different training sets, not from the method used to fit the emulator (PPE or VPPE). A quick ANOVA test shows that when we compare the sum of squares due to the fitting method to the total sum of squares, the proportion of variation explained by the method (PPE or VPPE) was only 0.03\% of the variation in RMSE across the 40 experiments (one PPE and one VPPE model for each of the 20 simulation experiments). The median RMSE across the 20 simulations for both PPE and VPPE was 0.00237/0.00236. Minimum and maximum values of the RMSE for both PPE and VPPE acrosss the 20 simulations were also 0.00211/0.00210 and 0.00278/0.00278, respectively. Thus, in terms of predictive performance, the PPE and VPPE approach to fitting range parameters are roughly equivalent.

To show VPPE's advantage in computational complexity on large training  datasets, we randomly split the dataset into 1696 training model runs, and 300 testing runs. Fitting the PPE emulator on a large number of model runs can take quite long, and Table \ref{tab:Joey_m} shows VPPE's runtimes and RMSE for the dataset for $m$ varying from 5 to 100.

\begin{table}[htbp]
\centering
\begin{tabular}{|c|*{6}{S|}}
\hline
$m$ & 5 & 15 & 25 & 50 & 75 & 100 \\
\hline
\textbf{VPPE} fit time (s) & 19.4 & 30.0 & 44.5 & 118 & 240 & 384\\
\hline
$\frac{\text{\textbf{VPPE} time}}{\text{\textbf{PPE} time}}$ & 0.04 & 0.06 & 0.09 & 0.24 & 0.49 & 0.79\\
\hline
\textbf{VPPE} RMSE $\times 10^{-4}$ & 6.37 & 6.64 & 6.76 & 6.82 & 6.82 & 6.86\\
\hline
\end{tabular}
\caption{Summary of the runtimes and RMSEs for VPPE for a fixed $n=1696$ and varying $m$ testing on $300$ runs. PPE took 488 seconds, and had a RMSE of $6.91 \times 10^{-4}$. The predict time is negligible.}
\label{tab:Joey_m}
\end{table}

We can see that as $m$ increases, the runtime increases for VPPE, but still remains low compared to PPE's runtime. While the RMSE is quite low compared to the range of model output values, it is interesting to see that increasing $m$ leads to {\it less} accurate emulators. All emulators identify saturated hydraulic conductivity ($K_s$) being the most influential input parameter, followed by the pore-size parameter ($\lambda$), and all emulators find porosity ($\theta$) to be markedly less influential. These parameters impact the shape of the output curve, suggesting that approximate likelihoods based on training curves with only the ``closest shapes" (small $m$) lead to more accurate predictions. Note that the only difference in the predictive GPs are estimates of their range parameters as shown in Table~\ref{tab:Joey_rp}; here indicating that porosity is even less influential on the resulting shape of the water content curve than it appears to be when we consider more (or all) neighbors.

\begin{figure}[htbp]
    \centering
    \includegraphics[width=0.75\linewidth]{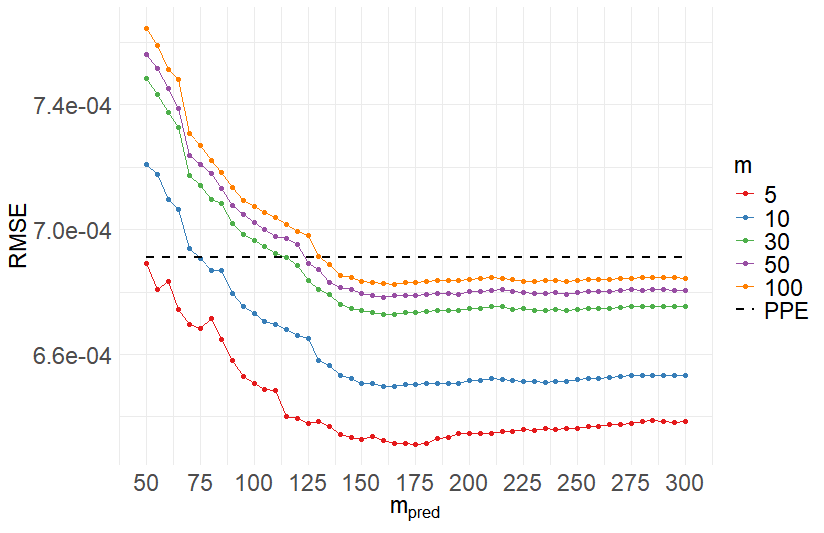}
    \caption{Plot of RMSE vs.~$m_{pred}$ the number of nearest neighbors used during prediction. Shown for different values of $m$, the number of nearest neighbors used while fitting the VPPE emulator. The black dashed line represents PPE's RMSE.}
    \label{fig:m_pred}
\end{figure}

To further explore this, we considered an experiment with VPPE's predictive mean built on only a subset of neighbors, $m_{pred}$.
In Fig.~\ref{fig:m_pred} we plot the predictive RMSE vs.~$m_{pred}$, the number of nearest neighbors used to predict. We can see that for all $m$ (number of neighbors to fit the VPPE), the optimal $m_{pred}$ lies between 150-175 out of the maximum $m_{pred}=300$ considered. The inclusion of ``too many" neighbors seems to slightly degrade predictions, showing Vecchia's advantage in its flexibility to move between local and global prediction modes.

\begin{table}[]
\centering
\begin{tabular}{|c|*{1}{J|}}
\hline
$m$ & $\hat{\boldsymbol{{\lambda}}}$ \\
\hline
5 & (10.4, 0.016, 0.29) \\
\hline
10 & (7.4, 0.012, 0.24) \\
\hline
30 & (5.7, 0.011, 0.21) \\
\hline
50 & (5.2, 0.011, 0.21) \\
\hline
100 & (4.7, 0.010, 0.21)\\
\hline
\end{tabular}
\caption{Estimates of the VPPE's normalized range parameters for $(\theta_s, K_s, \lambda)$ as $m$ increases from $m=5$ to $m=100$. The estimated normalized range parameters for PPE are $(4.5, 0.010, 0.21)$.} 
\label{tab:Joey_rp}
\end{table}

A common application of Vecchia approximations to simulator data with multidimensional (spatially varying) outputs is to treat space as an input. This has the effect of adding one (or two) input dimension and transforming the output to a scalar. This can increase the size of the training data dramatically, but of course, handling large training data is Vecchia's strong suit.

Recall in this  data set, the different output components ($k=100$) represent the soil depth ranging from 0.015 meters to 2.985 meters. Now we treat soil depth as an input parameter and consider the volumetric water content as a scalar response. We use the same training/testing split as earlier with 1696 training runs and 300 testing runs. Moving the output parameter to an input parameter transforms the data from 1696 model runs with three input dimensions and 100 output dimensions to 169,600 model runs with four input dimensions and one output dimension. Due to computational limitations, we sample 10\% of these model runs to fit the VRGaSP, reducing the number of training model runs to 16,960. When predicting on the remaining 300 testing runs, we predicted on all 100 output dimensions.

First, we train VRGaSP on varying $m$ from $m=5$ to $m=100$. The training times ranged from 99 seconds to 3652 seconds. We predict with a fixed $m_{pred}=150$ for all $m$. The results are shown in Table \ref{tab:Joey_m_k_1_v1}. 
\begin{table}[htbp]
\centering
\begin{tblr}{cells={valign=m,halign=c},
  hlines={},
  vlines={}}
$m$ & 5 & 15 & 25 & 50 & 75 & 100 \\
\textbf{VRGaSP} fit/predict time (s) & 99/112 & 181/116 & 353/116 & 1007/119 & 2163/109 & 3652/106\\
\textbf{VRGaSP} total time (s) & 211 & 297 & 469 & 1126 & 2272 & 3758\\
\textbf{VRGaSP} RMSE $\times 10^{-3}$ & 2.58 & 2.34 & 2.34 & 2.26 & 2.26 & 2.24
\end{tblr}
\caption{Summary of the runtimes and RMSEs for VRGaSP for a fixed $n=16,960$ and varying $m$ testing on $300$ runs. Recall from Table \ref{tab:Joey_m}, PPE took 488 seconds, and had a RMSE of $6.91 \times 10^{-4}$.}
\label{tab:Joey_m_k_1_v1}
\end{table}
We can see that compared to VPPE in Table \ref{tab:Joey_m}, the fit runtimes in this experiment are much longer. In fact, fitting VPPE took at the least 19 seconds and at the most 384 seconds, whereas fitting VRGaSP with space as an input took at the least 99 seconds and at the most 3652 seconds. Additionally, the RMSE is nearly a full order of magnitude larger, showing that VPPE provides faster and more accurate predictions to multidimensional outputs.

Another approach to transforming the multidimensional outputs is to use the 1696 training runs, and randomly sample one of the 100 output dimensions for each training run. This approach amounts roughly to extending the LHS design by a dimension to include the soil column depth. This transformation takes the simulator data from 1696 model runs with three input dimensions and 100 output dimensions to 1696 model runs with four input dimensions and one output dimension. With the number of training runs being significantly lower than the transformation just considered, we will predict using the full design rather than a set of nearest neighbors. 
\begin{table}[htbp]
\centering
\begin{tblr}{cells={valign=m,halign=c},
  hlines={},
  vlines={}}
$m$ & 5 & 15 & 25 & 50 & 75 & 100 \\
\textbf{VRGaSP} fit/predict times (s) & 9/22 & 18/20 & 32/20 & 99/20 & 204/19 & 346/19\\
\textbf{VRGaSP} total time (s) & 31 & 37 & 51 & 119 & 224 & 365\\
$\frac{\text{\textbf{VRGaSP} time}}{\text{\textbf{RGaSP} time}}$ & 0.02 & 0.03 & 0.04 & 0.09 & 0.18 & 0.29\\
\textbf{VRGaSP} RMSE $\times 10^{-3}$ & 6.46 & 5.79 & 5.67 & 5.66 & 5.67 & 5.58
\end{tblr}
\caption{Summary of the runtimes and RMSEs for VRGaSP for a fixed $n=1696$ and varying $m$ testing on $300$ runs. RGaSP took 1270 seconds, and had a RMSE of $5.59 \times 10^{-3}$.}
\label{tab:Joey_m_k_1_v2}
\end{table}
\renewcommand{\arraystretch}{1}
Results are shown in Table~\ref{tab:Joey_m_k_1_v2}, and as we expect, this experiment is fast but yields the least accurate predictions. We can see once again VRGaSP performing just as well as RGaSP at a fraction of the runtime. (Note, in the package \texttt{RobustGaSPV}, if one chooses to predict using nearest neighbors, the majority of the computational burden for prediction is in the prediction step as opposed to the fitting step. This accounts for discrepency in fit times between Tables~\ref{tab:Joey_m_k_1_v1} and \ref{tab:Joey_m_k_1_v2}.)
Overall, we see that VPPE is both faster than other options with improved or comparable predictive accuracy.

\subsection{Volcanic flow model}
Aluto volcano in Central Ethiopia has a complex topography, which heavily influences the dangerous flows composed of volcanic gas and particles following eruptions known as pyroclastic density currents (PDCs). In \cite{Tierz:etal:2024}, they combine the TITAN2D model \citep{Patra:etal:2005} for PDCs with the zero-censored Gaussian Process emulator \citep{Spill:etal:2023} to explore the topography's influence on PDC inundation patterns at reasonable computational cost. A key aleatoric input is the location where vents (that PDCs flow from) might open and multiple vent-opening models were considered in \cite{Tierz:etal:2024}. Exploring flow inundation from these possible vent locations necessitates a very large number of training runs. The output, max PDC inundation depth, also varies spatially. In short, this is an inherently large-$n$ and very large-$k$ data set.

The TITAN2D simulator implements a complex model of hyperbolic pdes of flow depth that conserve mass and momentum. These pdes are solved over digital elevation models and include flux sources of the flowing material. We consider the same computer model data set as \cite{Tierz:etal:2024} which consists of 4181 simulator runs. This data set has five input dimensions: vent radius (m), flux rate per unit area (m/s), bed friction angle (degrees), UTM Easting coordinate of the eruptive vent (m), and UTM Northing coordinate of the eruptive vent (m). 
The output data consists of the maximum flow depth for each pixel in a $1024 \times 1024$ grid, resulting in an output dimension of 1,048,576. We will restrict our data to the region of focus in \cite{Tierz:etal:2024}. This reduces our output dimension to 130,262. Additionally, we removed the simulator runs that yielded zero flow depth in the region of interest, reducing the number of runs from 4181 to 3864. 

Due to the $\mathcal{O}(n^2k)$ factor in PPE's computational complexity and the $\mathcal{O}(m^2k)$ factor in VPPE's, we chose to randomly sample 10\% of the output dimensions whilst finding the optimal range parameters for the emulators. In practice, we noticed that while this reduced the runtime significantly, it did not noticeably alter the final estimated range parameters. When estimating the range parameters for the emulator, we trained on all 3864 runs. After estimating the optimal range parameters, we randomly sampled a simulator run and fit the PPE and VPPE emulators with the other 3863 runs as training data and predicted on the sampled simulator run. While fitting the emulators, we did so on all 130,262 output dimensions.

\begin{figure}[htbp]
    \centering
    \includegraphics[width=0.95\textwidth]{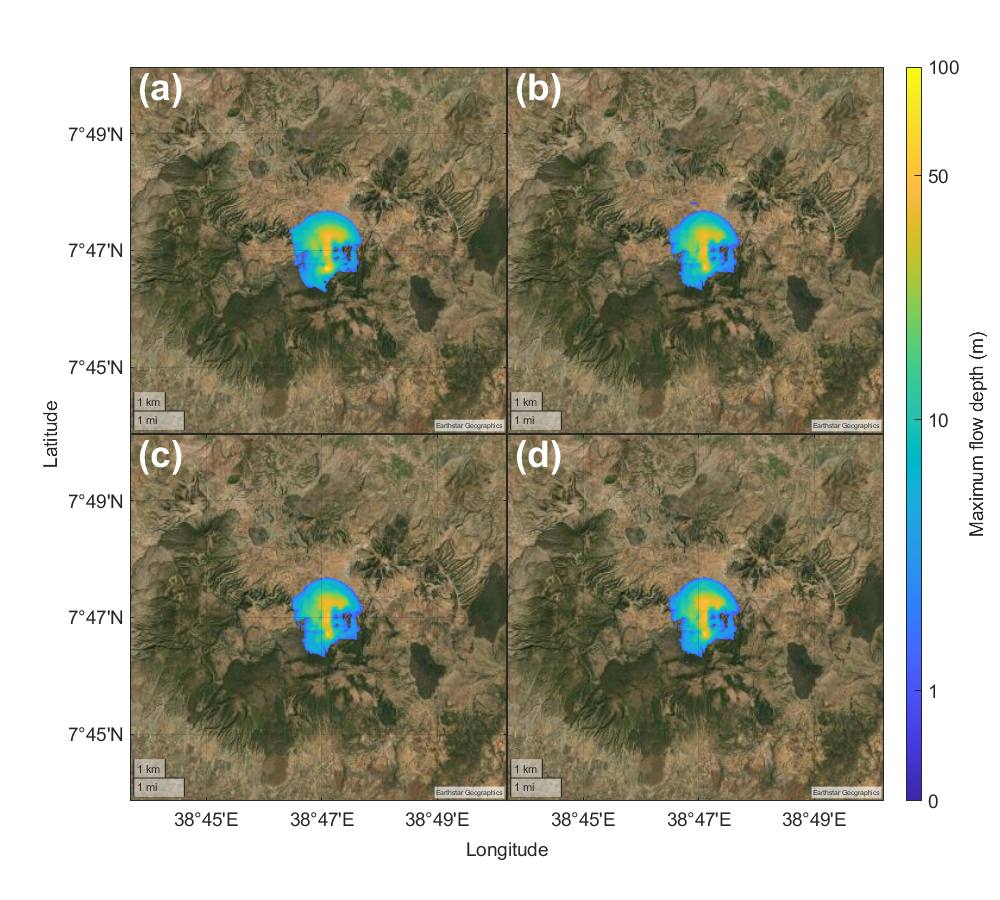}
    \caption{The simulator output (a) of a randomly sampled model run, compared to the predictions of emulators PPE (b), VPPE with $m_{pred} = 3863$ (c) and VPPE with $m_{pred}=200$ (d).}
    \label{fig:Aluto}
\end{figure}

For our VPPE emulator, we used $m=20$ nearest neighbors. The VPPE emulator took 3.34 hours to fit, while the PPE emulator took 23.66 hours to fit. The RMSE for PPE was 0.0952, whereas the RMSE for VPPE was 0.0959. In Fig.~\ref{fig:Aluto} we can see the predicted outputs for both PPE and VPPE compared to the exact output data from the simulator. The output data plotted on a log scale, and we have not plotted the pixels where the outputs or predicted outputs are below 0.5 meters.


We can see two main differences between the simulator's outputs and emulators' outputs. The first is that the maximum flow depth at the vent's location is a larger value in the simulator's output compared to the emulators' outputs. Additionally, the simulator's output values decrease faster as we are farther from the vent compared to the emulators' output values. Aside from that, the predicted outputs from both emulators are very accurate. VPPE is virtually indistinguishable from PPE at a fraction of the time needed to fit the emulator. 

Another aspect to consider is the number of nearest neighbors used while predicting. The default is to use all of the training data, but one has the option to select $m_{pred}$ number of nearest neighbors to the testing input parameters in order to predict faster. In Fig. \ref{fig:Aluto}, using $m_{pred} = 200$ yielded an RMSE of 0.0949; lower than both PPE and VPPE when using all of the training data to predict. Additionally, the runtime to predict was much faster, less than five seconds compared to over 17 minutes when predicting with the full training data. The global/local flexibility of the VPPE makes it very attractive for use in UQ studies relying on such complex geophysical simulators.

\section{Conclusion}\label{sec:Conclusion}

Here we introduced the new Vecchia parallel partial emulator developed for simulations with large training data sets and large-dimensional output. In lieu of relying on MLEs of the Vecchia approximate likelihood to find GP range parameters, we adopted the RGaSP framework \citep{Gu:Wang:Berg:2018, Gu:Palo:Berg:2019}. To begin, we derived the Vecchia marginal posterior, wherein we assumed a posterior with the Vecchia approximate likelihood and a reference prior of GP hyper parameters.  Subsequently, we derive the gradient of this Vecchia marginal posterior for use optimization schemes for finding the mode posterior. For scalar outputs, this resulted in the new VRGaSP approach, and for vector-valued outputs, the new VPPE.

Implementing the VPPE allows for parallel partial emulation with large $n$ that has comparable prediction results and drastically reduced runtime. 
To demonstate this, we applied the VPPE to three simulators -- a high-dimensional GP ``simulator", a hydrology simulator, and a volcanic flow simulator. In each case for moderate-to-large $n$, the VPPE performs on par with the PPE in terms of predictive accuracy, but with far lower computational cost. 

We see real value for the VPPE in large-$n$ training sets for geospatial simulators whose quantity of interest varies over large spatial domain. On its own, the PPE is an effective tool for UQ in such simulators because it handles spatial non-stationary implicitly by treating space as an output -- i.e.~each output component has its own scalar variance~\citep{Gu:Berg:2016}. Further as implemented, fitting a PPE utilizes the Robust GasP framework where the effects of uncertainty in the trend and variance parameters are quantified while estimating range parameters~\citep{Gu:Wang:Berg:2018, Gu:Palo:Berg:2019}. Lastly the global nature of the PPE is key to the surrogate maintaining conserved quantities (mass, momentum, etc.) that the simulator possesses~\citep{Gao:Pit:2024}. VPPE inherits all of these attractive properties from the PPE, but is computationally efficient in the large-$n$ case. Further, when the user values predictive speed over maintaining conserved quantities, the VPPE can simply be run in nearest-neighbor prediction mode as we saw with the volcanic flow simulator in Section~\ref{sec:Results}. 

\section*{Acknowledgments}
The authors would like to thank Pablo Tierz and Joey Lyon for providing us with the volcanic flow and hydrology simulation data, respectively.

\section{Appendix \label{sec:appendix}}
Here we will derive the Vecchia marignal posterior, beginning by multiplying the Vecchia likelihood, $L_m(\blambda, \boldsymbol{\beta}, \sigma^2 \mid \bx^\D, \by^\D)$, given in Eq.~\ref{eq:Vecchia_lik}, by the standard reference prior for $\bbeta$ and $\sigma^2$, $\pi^R(\boldsymbol{\beta}, \sigma^2) \propto \frac{1}{\sigma^2}$. We then integrate this product over $\boldsymbol{\beta}$ and $\sigma^2$. We will subsequently show that the Vecchia marginal posterior is equivalent to the marginal posterior in the RGaSP framework when the size of conditioning set is the rest of the training data. That is, $\mathcal{L}_m = \mathcal{L}$ when $m=n-1$.

We first integrate the posterior distribution over $\boldsymbol{\beta}$ and $\sigma^2$. The posterior is proportional to the product of the Vecchia likelihood approximation, $L_m(\mathbf{y})$ and the standard reference prior, proportional to $\frac{1}{\sigma^2}$ , i.e. we need to evaluate
\begin{align}
    \mathcal{L}_m &= \int_{\mathbb{R}^q \times (0,\infty)} L_m(\mathbf{y},\boldsymbol{\beta},\sigma^2) \cdot \frac{1}{\sigma^2} d\boldsymbol{\beta} d\sigma^2. \label{a1:integral_eq}
\end{align}
We will approach this by first integrating out $\boldsymbol{\beta}$, then integrating out $\sigma^2$. This will yield the Vecchia integrated likelihood $\mathcal{L}_m$. Then we prove that $\mathcal{L}_m$ and $\mathcal{L}$ are indeed equivalent when $m=n-1$, as one would expect.
This integral is proportional to
\begin{align}
    \mathcal{L}_m \propto \int_{\mathbb{R}^q \times (0,\infty)} (\sigma^2)^{-n/2} \left( \prod_{i=1}^n \omega_{im}^{-1/2} \right) \exp \left\{-\frac{1}{2\sigma^2} \sum_{i=1}^n \omega_{im}^{-1}e_{im}^2 \right\} \cdot \frac{1}{\sigma^2} d\boldsymbol{\beta} d\sigma^2 \label{a1:integral_eq2}
\end{align}

\subsection[Integrating out beta]{Integrating out $\boldsymbol{\beta}$ \label{sec:appendix_beta}}
We first focus on integrating with respect to $\boldsymbol{\beta}$ over $\mathbb{R}^q$ as shown in Eq. \ref{eq:vecchia_beta_integral}. To begin, it is instructive to rewrite $e_{im}$ as
\begin{align}
    e_{im} &= \varepsilon_{i} - \mathbf{r}_{im}^\top(\bR_{im}+\nu^2I)^{-1}\boldsymbol{\varepsilon}_{im},\\
    &= (y_i - \mathbf{h}(x_i)\boldsymbol{\beta}) - \mathbf{r}_{im}^\top(\bR_{im}+\nu^2I)^{-1}(\mathbf{y}_{im} - \mathbf{h}(x_{im})\boldsymbol{\beta}), \nonumber \\
    &= y_i - \bR_{im}^\top(\bR_{im}+\nu^2I)^{-1}\mathbf{y}_{im} - \mathbf{h}(x_{i})\boldsymbol{\beta} + \mathbf{r}_{im}^\top(\bR_{im}+\nu^2I)^{-1}\mathbf{h}(x_{im})\boldsymbol{\beta}. \nonumber
\end{align}
Denote $g_{im} = y_i - \mathbf{r}_{im}^\top(\bR_{im}+\nu^2I)^{-1}\mathbf{y}_{im}$, and $\tilde{\mathbf{h}}_{im} = \mathbf{h}(x_i) - \mathbf{r}_{im}^\top(\bR_{im}+\nu^2I)^{-1}\mathbf{h}(x_{im})$. Then the expression simplifies further as
\begin{align}
    e_{im} &= g_{im}- \mathbf{h}(x_{i})\boldsymbol{\beta} + \bR_{im}^\top(\bR_{im}+\nu^2I)^{-1}\mathbf{h}(x_{im})\boldsymbol{\beta},\\
    &= g_{im} - \tilde{\mathbf{h}}_{im}\boldsymbol{\beta}. \nonumber
\end{align}
The term $\omega_{im}^{-1}e_{im}^2$ in the argument of the exponential in Eq. \ref{a1:integral_eq2} can be rewritten as
\begin{align}
    \omega_{im}^{-1}e_{im}^2 &= e_{im} \omega_{im}^{-1} e_{im},\\
    &= (g_{im}-\tilde{\mathbf{h}}_{im}\boldsymbol{\beta})^\top\omega_{im}^{-1}(g_{im}-\tilde{\mathbf{h}}_{im}\boldsymbol{\beta}), \nonumber \\
    &= g_{im} \omega_{im}^{-1} g_{im} - \boldsymbol{\beta}^\top \tilde{\mathbf{h}}_{im}^\top\omega_{im}^{-1}g_{im} - g_{im}\omega_{im}^{-1}\tilde{\mathbf{h}}_{im}\boldsymbol{\beta} + \boldsymbol{\beta}^\top \tilde{\mathbf{h}}_{im}^\top\omega_{im}^{-1}\tilde{\mathbf{h}}_{im}\boldsymbol{\beta}. \nonumber
\end{align}
Since the expression $\omega_{im}^{-1}e_{im}^2$ is a scalar, the transpose of any of the individual terms above is equal to itself. This allows us to condense the expression further to be written as
\begin{align*}
    \omega_{im}^{-1}e_{im}^2 &= g_{im} \omega_{im}^{-1} g_{im} - 2\boldsymbol{\beta}^\top \tilde{\mathbf{h}}_{im}^\top\omega_{im}^{-1}g_{im} + \boldsymbol{\beta}^\top \tilde{\mathbf{h}}_{im}^\top\omega_{im}^{-1}\tilde{\mathbf{h}}_{im}\boldsymbol{\beta}.
\end{align*}
Now we include the summation in the exponent of Eq. \ref{a1:integral_eq2}. We get the following expression as
\begin{align*}
    \sum_{i=1}^n \omega_{im}^{-1}e_{im}^2 &= \sum_{i=1}^n g_{im} \omega_{im}^{-1} g_{im} - 2\boldsymbol{\beta}^\top \sum_{i=1}^n \tilde{\mathbf{h}}_{im}^\top\omega_{im}^{-1}g_{im} + \boldsymbol{\beta}^\top \left[\sum_{i=1}^n\tilde{\mathbf{h}}_{im}^\top\omega_{im}^{-1}\tilde{\mathbf{h}}_{im} \right]\boldsymbol{\beta}
\end{align*}
We can rearrange the expressions and complete the square for multivariate expressions using the formula $\boldsymbol{\beta}^\top A\boldsymbol{\beta} + 2\boldsymbol{\beta}^\top b + c = (\boldsymbol{\beta} + A^{-1}b)^\top A (\boldsymbol{\beta} + A^{-1}b) + c - b^\top A^{-1} b$. As a result,
\begin{align}
    \sum_{i=1}^n \omega_{im}^{-1}e_{im}^2 &= \boldsymbol{\beta}^\top \underbrace{\left[\sum_{i=1}^n\tilde{\mathbf{h}}_{im}^\top\omega_{im}^{-1}\tilde{\mathbf{h}}_{im} \right]}_{A}\boldsymbol{\beta} + 2\boldsymbol{\beta}^\top \underbrace{\left(-\sum_{i=1}^n \tilde{\mathbf{h}}_{im}^\top\omega_{im}^{-1}g_{im}\right)}_{b} + \underbrace{\sum_{i=1}^n g_{im}^\top \omega_{im}^{-1} g_{im}}_{c},\\
    &= (\boldsymbol{\beta} - \boldsymbol{\mu})^\top \tilde{\mathrm{\Sigma}}(\boldsymbol{\beta}-\boldsymbol{\mu}) + \tilde{S}^2, \nonumber
\end{align}
where
\begin{align}
    \tilde{\mathrm{\Sigma}} &= \sum_{i=1}^n\tilde{\mathbf{h}}_{im}^\top\omega_{im}^{-1}\tilde{\mathbf{h}}_{im}, \\
    \boldsymbol{\mu} &= \tilde{\mathrm{\Sigma}}^{-1} \left[ \sum_{i=1}^n \tilde{\mathbf{h}}_{im}^\top\omega_{im}^{-1}g_{im}\right], \nonumber \\
    \tilde{S}^2 &= \sum_{i=1}^n g_{im}^\top \omega_{im}^{-1} g_{im} - \left[ \sum_{i=1}^n \tilde{\mathbf{h}}_{im}^\top\omega_{im}^{-1}g_{im} \right]^\top \tilde{\mathrm{\Sigma}}^{-1}\left[ \sum_{i=1}^n \tilde{\mathbf{h}}_{im}^\top\omega_{im}^{-1}g_{im} \right], \nonumber \\
    &= \sum_{i=1}^n g_{im}^\top \omega_{im}^{-1} g_{im} - \boldsymbol{\mu}^\top\tilde{\mathrm{\Sigma}} \boldsymbol{\mu}. \nonumber
\end{align}
Here, although the $q \times q$ matrix $\tilde{\mathrm{\Sigma}}$ is not denoted by an inversion symbol, it is a combination of inverse conditional variances $\omega_{im}^{-1}$ through $\tilde{\mathbf{h}}_{im}$. Additionally, we can note that plugging in the profile likelihood estimate $\bbeta_m$ from Eq. \ref{eq:vecchia_log_lik_profiled} reveals that $\tilde{S}^2$ is equivalent to $n \sigma^2_m$ from Eq. \ref{eq:vecchia_log_lik_profiled} as shown here,
\begin{align}\label{eq:sigmisS}
    n \sigma^2_m &= \sum_{i=1}^n \omega_{im}^{-1} \tilde{e}_{im}^2, \\
    &= \sum_{i=1}^n g_{im} \omega_{im}^{-1} g_{im} - 2\bbeta_m^\top \sum_{i=1}^n \tilde{\mathbf{h}}_{im}^\top\omega_{im}^{-1}g_{im} + \bbeta_m^\top \left[\sum_{i=1}^n\tilde{\mathbf{h}}_{im}^\top\omega_{im}^{-1}\tilde{\mathbf{h}}_{im} \right]\bbeta_m, \nonumber \\
    &= \sum_{i=1}^n g_{im} \omega_{im}^{-1} g_{im} - \left[ \sum_{i=1}^n \omega_{im}^{-1}\tilde{\mathbf{h}}_{im}^\top g_{im}\right]^\top \left[\sum_{i=1}^n \omega_{im}^{-1}\tilde{\mathbf{h}}_{im} \tilde{\mathbf{h}}_{im}^\top \right]^{-\top} \left[\sum_{i=1}^n \tilde{\mathbf{h}}_{im}^\top\omega_{im}^{-1}g_{im} \right], \nonumber \\
    &= \sum_{i=1}^n g_{im} \omega_{im}^{-1} g_{im} - \left[ \sum_{i=1}^n \omega_{im}^{-1}\tilde{\mathbf{h}}_{im}^\top g_{im}\right]^\top \left[\sum_{i=1}^n \omega_{im}^{-1}\tilde{\mathbf{h}}_{im}^\top \tilde{\mathbf{h}}_{im} \right]^{-1} \left[\sum_{i=1}^n \tilde{\mathbf{h}}_{im}^\top\omega_{im}^{-1}g_{im} \right], \nonumber \\
    &= \tilde{S}^2. \nonumber 
\end{align}
Note that the last equality follows the definition of $\tilde{S}^2$ given in Eq. \ref{eq:Ssqrd}. Now we can write the full exponent in Eq. \ref{a1:integral_eq2} in a form that allows us to integrate over $\boldsymbol{\beta}$. The exponential term can be rewritten as
\begin{align}
    \exp \left\{-\frac{1}{2\sigma^2} \sum_{i=1}^n \omega_{im}^{-1}e_{im}^2 \right\} &= \exp \left\{ -\frac{1}{2\sigma^2} \left[(\boldsymbol{\beta} - \boldsymbol{\mu})^\top \tilde{\mathrm{\Sigma}}(\boldsymbol{\beta}-\boldsymbol{\mu}) + \tilde{S}^2 \right]\right\},\\
    &= \exp \left\{-\frac{\tilde{S}^2}{2\sigma^2}\right\}\exp \left\{ -\frac{1}{2\sigma^2} \left[(\boldsymbol{\beta} - \boldsymbol{\mu})^\top \tilde{\mathrm{\Sigma}}(\boldsymbol{\beta}-\boldsymbol{\mu}) \right]\right\}. \nonumber \\
\end{align}
The integral in Eq. \ref{a1:integral_eq2} simplifies to
\begin{align}
    \int_0^\infty (\sigma^2)^{-n/2-1} \left( \prod_{i=1}^n \omega_{im}^{-1/2} \right) \exp \left\{-\frac{\tilde{S}^2}{2\sigma^2}\right\} \Bigg[\int_{\mathbb{R}^q} \exp \left\{ -\frac{1}{2\sigma^2} \left[(\boldsymbol{\beta} - \boldsymbol{\mu})^\top \tilde{\mathrm{\Sigma}}(\boldsymbol{\beta}-\boldsymbol{\mu}) \right]\right\} d\boldsymbol{\beta} \Bigg] d\sigma^2. \label{a1:separable_int}
\end{align}
Note that the $\boldsymbol{\beta}$ integral is very similar to the multivariate normal distribution integral. We can rewrite the exponent to get it in the correct form. Additionally, we can remove $\boldsymbol{\mu}$ since a shift in $\mathbb{R}^q$ does not change the resulting integral, which is equal to
\begin{align*}
    \int_{\mathbb{R}^q} \exp \left\{ -\frac{1}{2\sigma^2} \left[\boldsymbol{\beta}^\top \tilde{\mathrm{\Sigma}}\boldsymbol{\beta} \right]\right\} d\boldsymbol{\beta} &= \int_{\mathbb{R}^q} \exp \left\{ -\frac{1}{2} \left[\boldsymbol{\beta}^\top \left\{(\sigma^2)^{-1}\tilde{\mathrm{\Sigma}}\right\}\boldsymbol{\beta}\right]\right\} d\boldsymbol{\beta}.
\end{align*}
We can see that this is close to being a normal distribution with covariance matrix $\left\{(\sigma^2)^{-1} \tilde{\mathrm{\Sigma}}\right\}^{-1}$. Multiplying and dividing by a normalizing constant (with respect to $\boldsymbol{\beta}$) allows us to evaluate this integral,
\begin{align}
    \int_{\mathbb{R}^q} \exp \left\{ -\frac{1}{2} \left[\boldsymbol{\beta}^\top \left\{(\sigma^2)^{-1}\tilde{\mathrm{\Sigma}}\right\}\boldsymbol{\beta}\right]\right\} d\boldsymbol{\beta} &= \frac{\sqrt{(2\pi)^q |(\sigma^2)^{-1} \tilde{\mathrm{\Sigma}}|^{-1}}}{\sqrt{(2\pi)^q |(\sigma^2)^{-1} \tilde{\mathrm{\Sigma}}|^{-1}}} \int_{\mathbb{R}^q} \exp \left\{ -\frac{1}{2} \left[\boldsymbol{\beta}^\top \left\{(\sigma^2)^{-1}\tilde{\mathrm{\Sigma}}\right\}\boldsymbol{\beta}\right]\right\} d\boldsymbol{\beta}, \\
    &= \sqrt{(2\pi)^q |(\sigma^2)^{-1} \tilde{\mathrm{\Sigma}}|^{-1}}, \nonumber \\
    &\propto (\sigma^2)^{q/2} |\tilde{\mathrm{\Sigma}}|^{-1/2}, \nonumber 
\end{align}
where $|\cdot|$ represent the determinant operation. Now we can insert this result into Eq. \ref{a1:separable_int} to get the expression,
\begin{align}
    \mathcal{L}_m \propto \int_0^\infty (\sigma^2)^{-n/2-1} \left( \prod_{i=1}^n \omega_{im}^{-1/2} \right) \exp \left\{-\frac{\tilde{S}^2}{2\sigma^2}\right\} \left[ (\sigma^2)^{q/2} |\tilde{\mathrm{\Sigma}}|^{-1/2} \right]d\sigma^2. \label{a1:beta_int_done}
\end{align}
\subsection[Integrating out sigma2]{Integrating out $\sigma^2$ \label{sec:appendix_sigma2}}
Now that we have computed the integral over $\boldsymbol{\beta}$, we can compute the integral over $\sigma^2$ as well. First, we simplify the expression to
\begin{align}
      \mathcal{L}_m \propto \left( \prod_{i=1}^n \omega_{im}^{-1/2} \right) |\tilde{\mathrm{\Sigma}}|^{-1/2} \int_0^\infty (\sigma^2)^{-\frac{(n-q)}{2}-1}  \exp \left\{-\frac{\tilde{S}^2}{2\sigma^2}\right\} d\sigma^2. \label{a1:beta_int_done2}
\end{align}
Now we can evaluate the integral over $\sigma^2$ as shown in Eq. \ref{eq:vecchia_sigma2_integral} to be
\begin{align}
       I_{\sigma^2} = \int_0^\infty (\sigma^2)^{-\frac{(n-q)}{2}-1}  \exp \left\{-\frac{\tilde{S}^2}{2\sigma^2}\right\} d\sigma^2. \label{a1:sig2_int} 
\end{align}
First, we perform a change in variables where $\tau = \frac{\tilde{S}^2}{2\sigma^2}$, or $\sigma^2 = \frac{\tilde{S}^2}{2\tau}$. As a result, $d\tau = -\frac{\tilde{S}^2}{2(\sigma^2)^2} d\sigma^2$. Rearranging to solve for $d\sigma^2$ gives us $d\sigma^2 = -\frac{\tilde{S}^2}{2\tau^2}$. Our integral bounds are reversed and Eq. \ref{a1:sig2_int} becomes
\begin{align}
    I_{\sigma^2} &= \int_{\infty}^0 \left(\frac{\tilde{S}^2}{2\tau}\right)^{-\frac{(n-q)}{2}-1} \exp\left\{ -\tau\right\} \left(-\frac{\tilde{S}^2}{2\tau^2}\right) d\tau,\\
    &= -\int_\infty^0 \left(\frac{\tilde{S}^2}{2\tau}\right)^{-\frac{(n-q)}{2}} \cdot \frac{e^{-\tau}}{\tau} d\tau, \nonumber \\
    &= \left( \frac{\tilde{S}^2}{2}\right)^{-\frac{n-q}{2}} \int_0^\infty \tau^{\frac{(n-q)}{2}-1} e^{-\tau} d\tau.  \nonumber 
\end{align}
We can see that the resulting integral is the definition of the gamma function. Our integral becomes
\begin{align*}
    I_{\sigma^2} =\left( \frac{\tilde{S}^2}{2}\right)^{-\frac{n-q}{2}} \mathrm{\Gamma}\left(\frac{n-q}{2}\right),
\end{align*}
and we can insert this into Eq. \ref{a1:beta_int_done2} to obtain the result for our double integral as
\begin{align}
    \mathcal{L}_m &\propto \left( \prod_{i=1}^n \omega_{im}^{-1/2} \right) |\tilde{\mathrm{\Sigma}}|^{-1/2} \left( \frac{\tilde{S}^2}{2}\right)^{-\frac{n-q}{2}} \mathrm{\Gamma}\left(\frac{n-q}{2}\right),\\
    &\propto \left( \prod_{i=1}^n \omega_{im}^{-1/2} \right) |\tilde{\mathrm{\Sigma}}|^{-1/2} \left(\tilde{S}^2\right)^{-\frac{n-q}{2}}. \nonumber 
\end{align}
Therefore, our marginalized likelihood function for the Vecchia approximation as shown in Eq. \ref{eq:result} is 
\begin{align}
    \mathcal{L}_m(\mathbf{y}) &\propto \left( \prod_{i=1}^n \omega_{im}^{-1/2} \right) |\tilde{\mathrm{\Sigma}}|^{-1/2} \left(\tilde{S}^2\right)^{-\frac{n-q}{2}}, \label{a1:result}
\end{align}
where we recall
\begin{align}
    \tilde{\mathrm{\Sigma}} &= \sum_{i=1}^n\tilde{\mathbf{h}}_{im}^\top\omega_{im}^{-1}\tilde{\mathbf{h}}_{im},\\
    \tilde{S}^2 &= \sum_{i=1}^n g_{im}^\top \omega_{im}^{-1} g_{im} - \left[ \sum_{i=1}^n \tilde{\mathbf{h}}_{im}^\top\omega_{im}^{-1}g_{im} \right]^\top \left[\sum_{i=1}^n\tilde{\mathbf{h}}_{im}^\top\omega_{im}^{-1}\tilde{\mathbf{h}}_{im} \right]\left[ \sum_{i=1}^n \tilde{\mathbf{h}}_{im}^\top\omega_{im}^{-1}g_{im} \right]. \nonumber 
\end{align}
\subsection[Proving equivalency]{Showing $\mathcal{L}_m = \mathcal{L}$ when $m=n-1$ \label{sec:appendix_equivalence}}
If we compare Eq. \ref{a1:result} to Eq. \ref{eq:likelihood}, we see corresponding terms in the equations. Indeed, the two are equivalent when $m=n-1$, as we would expect. We will show that $$\mathcal{L}_m(\mathbf{y}) \propto \left( \prod_{i=1}^n \omega_{im}^{-1/2} \right) |\tilde{\mathrm{\Sigma}}|^{-1/2} \left(\tilde{S}^2\right)^{-\frac{n-q}{2}}$$ is equivalent to $$\mathcal{L}(\boldsymbol{\lambda} | \mathbf{y}^\mathcal{D}) \propto |\tilde{\mathbf{R}}|^{-1/2}|\htxD \tilde{\mathbf{R}}^{-1} \hxD|^{-1/2} \left(S^2 \right)^{-\left(\frac{n-q}{2}\right)}$$ when $m = n-1$. Here $\tilde{\mathbf{R}} = \mathbf{R} + \nu^2 I$ to account for noise.

Clearly, $\prod_{i=1}^n \omega_{im}^{-1/2} = |\tilde{\mathbf{R}}|$ when $m=n-1$ from \cite{Vecchia:1988}. It remains to show that when $m=n-1$, $\tilde{\mathrm{\Sigma}}$ is equivalent to $\mathbf{h}^\top(\bx) \tilde{\mathbf{R}}^{-1}\mathbf{h}(\bx)$, and $\tilde{S}^2 = S^2$ from Eq. \ref{eq:likelihood}.

We are given an $n \times n$ covariance matrix $\tilde{\mathbf{R}}$ obtained from design $\bx^\mathcal{D}$ with dimensions $n \times p$, where each design run $\bx_i$ is a $p$-dimensional row vector. Take vectors $\mathbf{a}, \mathbf{b}$ in $\mathbb{R}^n$. Here $a_i$ is the $i^{th}$ element of $\mathbf{a}$ and $a_{im}$ is the subvector of $\mathbf{a}$ corresponding to the nearest neighbors to $x_i$. 
We can show the following expression is true.
\begin{align}
    \mathbf{a}^\top \tilde{\mathbf{R}}^{-1} \mathbf{b} &= \sum_{i=1}^n \left(a_i-\boldsymbol{\rho}_{i,im} \boldsymbol{\rho}_{im}^{-1} a_{im} \right)^\top \mathrm{\Sigma}_{i|im}^{-1} \left(b_i-\boldsymbol{\rho}_{i,im} \boldsymbol{\rho}_{im}^{-1} b_{im}\right), \label{a1:quadratic_form}
\end{align}
where $\boldsymbol{\rho}_{i,im}$ and $\boldsymbol{\rho}_{im}$ are the correlations between $\bx_{i}$ and $\bx_{im}$, and $\bx_{im}$ and $\bx_{im}$, respectively. $\mathrm{\Sigma}_{i|im}$ is the conditional variance of $\bx_i$ conditioned on $\bx_{im}$. Additionally, by the definitions stated in \cite{Vecchia:1988}, we can rewrite Eq. \ref{a1:quadratic_form} in our notation to be
\begin{align}
    \mathbf{a}^\top \tilde{\mathbf{R}}^{-1} \mathbf{b} &= \sum_{i=1}^n \left(a_i-\mathbf{r}_{im}^\top \left(\bR_{im}+\nu^2 I \right)^{-1} \mathbf{a}_{im} \right)^\top \omega_{im}^{-1} \left(b_i-\mathbf{r}_{im}^\top \left(\bR_{im}+\nu^2 I \right)^{-1} \mathbf{b}_{im}\right). \label{a1:quadratic_form2}
\end{align}
It is easy to see that the expression $\mathbf{r}_{im}^\top \left(\bR_{im}+\nu^2 I \right)^{-1} \mathbf{a}_{im}$ is equivalent to the linear regression estimate of $a_i$ conditioned on $\mathbf{a}_{im}$ \citep{Sant:Will:2003}. We can rewrite the LHS of Eq. \ref{a1:quadratic_form2} by taking the Cholesky decomposition of $\tilde{\mathbf{R}}^{-1}=\mathbf{L}^\top \mathbf{L}$ as
\begin{align}
    \mathbf{a}^\top \tilde{\mathbf{R}}^{-1} \mathbf{b} &= \mathbf{a}^\top \left(\mathbf{L}^\top \mathbf{L} \right) \mathbf{b}, \\
    &= \left(\mathbf{a}^\top \mathbf{L}^{\top} \right) \left(\mathbf{L} \mathbf{b} \right),  \nonumber \\
    &= \left(\mathbf{L} \mathbf{a}\right)^{\top} \left(\mathbf{L} \mathbf{b} \right),  \nonumber \\
    &= \mathbf{z_a}^\top \mathbf{z_b},  \nonumber \\
    &= \sum_{i=1}^n \left(\mathbf{z_a}\right)_i \left(\mathbf{z_b}\right)_i, \nonumber 
\end{align}
where $\mathbf{z_a} = \mathbf{L} \mathbf{a}$ and $\mathbf{z_b} = \mathbf{L} \mathbf{b}$. If we define $a_t = \sum_{i=1}^{t-1} \phi_{ti}a_i + \epsilon_i$, where $\phi_{ti}$ are the regression parameters of $a_t$ conditioned on $a_1,\dots,a_{t-1}$, we know that the precision matrix $\mathrm{\Sigma}^{-1}$ has the modified Cholesky decomposition \citep{Pourahmadi:1999, Pourahmadi:2000, Dai:Pan:Liang:2023}
\begin{align}
    \mathrm{\Sigma}^{-1} &= \mathbf{T}' \mathbf{D}^{-1} \mathbf{T},
\end{align}
with $\mathbf{T}$ defined as
\begin{align}
    \mathbf{T} &= \begin{bmatrix}
        1 & 0 & 0 & \cdots & 0 \\
        -\phi_{21} & 1 & 0 & \cdots & 0\\
        -\phi_{31} & -\phi_{32} & 1 & \cdots & 0\\
        \vdots & \vdots & \vdots & \ddots & \vdots \\
        -\phi_{n1} & -\phi_{n2} & -\phi_{n3} & \cdots & 1
    \end{bmatrix},
\end{align}
and $\mathbf{D}$ is the diagonal matrix of conditional variances $(\omega_1, \omega_2, \dots, \omega_n)$. Then $\mathbf{L}^\top = \mathbf{T}'\mathbf{D}^{-1/2}$, and $\mathbf{L} = \mathbf{D}^{-1/2}\mathbf{T}$ has the form 
\begin{align}
    \mathbf{L} &= \begin{bmatrix}
        \omega_{1}^{-1/2} & 0 & 0 & \cdots & 0\\
        -\phi_{21} \omega_{2}^{-1/2} & \omega_{2}^{-1/2} & 0 & \cdots & 0 \\
        -\phi_{31}\omega_{3}^{-1/2} & -\phi_{32}\omega_{3}^{-1/2} & \omega_{3}^{-1/2} & \cdots & 0\\
        \vdots & \vdots & \vdots & \ddots & \vdots \\
        -\phi_{n1}\omega_{n}^{-1/2} & -\phi_{n2}\omega_{n}^{-1/2} & -\phi_{n3}\omega_{n}^{-1/2} & \cdots & \omega_{n}^{-1/2}
    \end{bmatrix}. \label{a1:L_formula}
\end{align}
Since $\mathbf{z}_a = \mathbf{L}\mathbf{a}$, we can see the expression $\mathbf{z_a}$ is equal to
\begin{align}
    \mathbf{z_a} &= \begin{bmatrix}
        a_1 \omega_{1}^{-1/2} \\
        \left(a_2 - \phi_{21}a_1\right) \omega_{2}^{-1/2} \\
        \left(a_3 - \phi_{31}a_1 - \phi_{32}a_2\right) \omega_{3}^{-1/2} \\
        \vdots\\
        \left(a_n - \sum_{i=1}^{n-1} \phi_{ni}a_i\right)\omega_{n}^{-1/2}
    \end{bmatrix}. \label{a1:z_a}
\end{align}
Since by definition, $\sum_{i=1}^{t-1} \phi_{ti}a_i$ is the regression estimate of $a_t$, we can see that $\left(\mathbf{z_a}\right)_i = \left(a_i-\mathbf{r}_{im}^\top \left(\bR_{im}+\nu^2 I \right)^{-1} \mathbf{a}_{im} \right)\omega_{im}^{-1/2}$. The logic for $\mathbf{z_b}$ is equivalent and so we have proven Eq. \ref{a1:quadratic_form}.

Since this was shown for arbitrary vectors $\mathbf{a}$ and $\mathbf{b}$, it follows that when $m=n-1$, $\tilde{\mathrm{\Sigma}} = \mathbf{h}^\top(\bx) \tilde{\mathbf{R}}^{-1}\mathbf{h}(\bx)$ and $\tilde{S}^2 = S^2$. For element $(s,j)$ of $\tilde{\mathrm{\Sigma}}$,
\begin{align}
    \tilde{\mathrm{\Sigma}}_{sj} &= \sum_{i=1}^n (\tilde{\mathbf{h}}_{im})_s^\top\omega_{im}^{-1}(\tilde{\mathbf{h}}_{im})_j, \\
    &= \sum_{i=1}^n \left(\mathbf{r}_{im}^\top(\bR_{im}+\nu^2I)^{-1}(\tilde{\mathbf{h}}_{im})_s - \mathbf{h}(x_i)\right)^\top\omega_{im}^{-1}\left(\mathbf{r}_{im}^\top(\bR_{im}+\nu^2I)^{-1}(\tilde{\mathbf{h}}_{im})_j - \mathbf{h}(x_i)\right), \nonumber  \\
    &= \sum_{i=1}^n \left(\mathbf{h}(x_i)-\mathbf{r}_{im}^\top(\bR_{im}+\nu^2I)^{-1}(\tilde{\mathbf{h}}_{im})_s\right)^\top\omega_{im}^{-1}\left(\mathbf{h}(x_i)-\mathbf{r}_{im}^\top(\bR_{im}+\nu^2I)^{-1}(\tilde{\mathbf{h}}_{im})_j\right), \nonumber \\
    &= (\mathbf{h}^\top(\bx))_s \tilde{\mathbf{R}}^{-1}(\mathbf{h}(\bx))_j, \nonumber 
\end{align}
where $(\tilde{\mathbf{h}}_{im})_s$ is the $s^{th}$ row vector of $\tilde{\mathbf{h}}_{im}$. Since this holds for each element of $\mathrm{\tilde{\Sigma}}$, it must hold for the full matrix. Therefore, $\tilde{\mathrm{\Sigma}} = \mathbf{h}^\top(\bx) \tilde{\mathbf{R}}^{-1}\mathbf{h}(\bx)$.

For $\tilde{S}^2$ we must apply our result to a few terms. First note that
\begin{align}
    \sum_{i=1}^n g_{im} \omega_{im}^{-1} g_{im} &= \sum_{i=1}^n \left(y_i-\mathbf{r}_{im}^\top(\bR_{im}+\nu^2I)^{-1}\mathbf{y}_{im}\right)^\top\omega_{im}^{-1}\left(y_i-\mathbf{r}_{im}^\top(\bR_{im}+\nu^2I)^{-1}\mathbf{y}_{im}\right),\\
    &= \mathbf{y}^\top \tilde{\mathbf{R}}^{-1}\mathbf{y}, \nonumber 
\end{align}
and
\begin{align}
    \sum_{i=1}^n \tilde{\mathbf{h}}_{im}^\top \omega_{im}^{-1} g_{im} &= \sum_{i=1}^n \left(\mathbf{h}(x_i)-\mathbf{r}_{im}^\top(\bR_{im}+\nu^2I)^{-1}\mathbf{h}(x_{im})\right)^\top\omega_{im}^{-1}\left(y_i-\mathbf{r}_{im}^\top(\bR_{im}+\nu^2I)^{-1}\mathbf{y}_{im}\right),\\
    &= \mathbf{h}(\bx)^\top \tilde{\mathbf{R}}^{-1}\mathbf{y}, \nonumber 
\end{align}
therefore,
\begin{align}
    \tilde{S}^2 &= \sum_{i=1}^n g_{im}^\top \omega_{im}^{-1} g_{im} - \left[ \sum_{i=1}^n \tilde{\mathbf{h}}_{im}^\top\omega_{im}^{-1}g_{im} \right]^\top \tilde{\mathrm{\Sigma}}^{-1}\left[ \sum_{i=1}^n \tilde{\mathbf{h}}_{im}^\top\omega_{im}^{-1}g_{im} \right],\\
    &= \mathbf{y}^\top\tilde{\mathbf{R}}^{-1}\mathbf{y} - \mathbf{y}^\top \tilde{\mathbf{R}}^{-1} \mathbf{h}(\bx) \left\{\mathbf{h}^\top(\bx) \tilde{\mathbf{R}}^{-1}\mathbf{h}^\top(\bx)\right\}^{-1} \mathbf{h}^\top(\bx) \tilde{\mathbf{R}}^{-1} \mathbf{y} \nonumber, \\
    &= S^2, \nonumber 
\end{align}
and we have shown that the integrated likelihood for the Vecchia approximation is equivalent to the full integrated likelihood when $m=n-1$.\\

\bibliography{bib}

\begin{thebibliography}{}

\bibitem[Balcan et~al., 2010]{Balcan:etal:2010}
Balcan, D., Gon{\c{c}}alves, B., Hu, H., Ramasco, J.~J., Colizza, V., and
  Vespignani, A. (2010).
\newblock Modeling the spatial spread of infectious diseases: The global
  epidemic and mobility computational model.
\newblock {\em Journal of computational science}, 1(3):132--145.

\bibitem[Carnell, 2024]{Carnell:2024}
Carnell, R. (2024).
\newblock {\em lhs: Latin Hypercube Samples}.
\newblock R package version 1.2.0.

\bibitem[Currin et~al., 1988]{Curr:etal:1988}
Currin, C., Mitchell, T., Morris, M.~D., and Ylvisaker, D. (1988).
\newblock A {B}ayesian approach to the design and analysis of computer
  experiments.
\newblock Technical report, Oak Ridge National Laboratory, Oak Ridge, TN (USA).

\bibitem[Dai et~al., 2022]{Dai:Pan:Liang:2023}
Dai, D., Pan, J., and and, Y.~L. (2022).
\newblock Regularized estimation of the mahalanobis distance based on modified
  cholesky decomposition.
\newblock {\em Communications in Statistics: Case Studies, Data Analysis and
  Applications}, 8(4):559--573.

\bibitem[Gao and Pitman, 2024]{Gao:Pit:2024}
Gao, Y. and Pitman, E.~B. (2024).
\newblock Parallel partial emulation in applications.
\newblock {\em International Journal for Uncertainty Quantification},
  14(6):1--15.

\bibitem[Gramacy, 2016]{Gramacy:2016}
Gramacy, R.~B. (2016).
\newblock {laGP}: Large-scale spatial modeling via local approximate gaussian
  processes in {R}.
\newblock {\em Journal of Statistical Software}, 72(1):1--46.

\bibitem[Gramacy, 2020]{Gram:2020}
Gramacy, R.~B. (2020).
\newblock {\em Surrogates: Gaussian process modeling, design, and optimization
  for the applied sciences}.
\newblock Chapman and Hall/CRC.

\bibitem[Gramacy and Apley, 2015]{Gram:Aple:2015}
Gramacy, R.~B. and Apley, D.~W. (2015).
\newblock Local gaussian process approximation for large computer experiments.
\newblock {\em Journal of Computational and Graphical Statistics},
  24(2):561--578.

\bibitem[Gu and Berger, 2016]{Gu:Berg:2016}
Gu, M. and Berger, J.~O. (2016).
\newblock Parallel partial gaussian process emulation for computer models with
  massive output.
\newblock {\em The Annals of Applied Statistics}, 10(3):1317--1347.

\bibitem[Gu et~al., 2019]{Gu:Palo:Berg:2019}
Gu, M., Palomo, J., and Berger, J.~O. (2019).
\newblock Robustgasp: Robust {G}aussian stochastic process emulation in r.
\newblock {\em The R Journal}, 11(1):112--136.

\bibitem[Gu et~al., 2018]{Gu:Wang:Berg:2018}
Gu, M., Wang, X., and Berger, J.~O. (2018).
\newblock Robust {G}aussian stochastic process emulation.
\newblock {\em The Annals of Statistics}, 46(6A):3038--3066.

\bibitem[Guinness, 2018]{Guinness:2018}
Guinness, J. (2018).
\newblock Permutation and grouping methods for sharpening gaussian process
  approximations.
\newblock {\em Technometrics}, 60(4):415--429.
\newblock PMID: 31447491.

\bibitem[Guinness, 2019]{Guinness:2019}
Guinness, J. (2019).
\newblock Gaussian process learning via fisher scoring of vecchia's
  approximation.

\bibitem[Guinness, 2021]{Guiness:2021}
Guinness, J. (2021).
\newblock Gaussian process learning via fisher scoring of vecchia’s
  approximation.
\newblock {\em Statistics and Computing}, 31(3).

\bibitem[Katzfuss et~al., 2022]{Katz:Guin:Lawr:2022}
Katzfuss, M., Guinness, J., and Lawrence, E. (2022).
\newblock Scaled vecchia approximation for fast computer-model emulation.
\newblock {\em SIAM/ASA Journal on Uncertainty Quantification}, 10(2):537--554.

\bibitem[Kennedy and O'Hagan, 2001]{Kenn:OHag:2001}
Kennedy, M.~C. and O'Hagan, A. (2001).
\newblock Bayesian calibration of computer models.
\newblock {\em Journal of the Royal Statistical Society: Series B (Statistical
  Methodology)}, 63(3):425--464.

\bibitem[Lawrence et~al., 2024]{Lawrence:etal:2024}
Lawrence, I.~R., Ridout, A.~L., Shepherd, A., and Tilling, R. (2024).
\newblock A simulation of snow on antarctic sea ice based on satellite data and
  climate reanalyses.
\newblock {\em Journal of Geophysical Research: Oceans}, 129(1):e2022JC019002.
\newblock e2022JC019002 2022JC019002.

\bibitem[Liu and Nocedal, 1989]{Liu:Nocedal:1989}
Liu, D.~C. and Nocedal, J. (1989).
\newblock On the limited memory bfgs method for large scale optimization.
\newblock {\em Mathematical programming}, 45(1):503--528.

\bibitem[Marchildon and Zingg, 2023]{Marc:Zing:2023}
Marchildon, A.~L. and Zingg, D.~W. (2023).
\newblock A non-intrusive solution to the ill-conditioning problem of the
  gradient-enhanced gaussian covariance matrix for gaussian processes.
\newblock {\em J. Sci. Comput.}, 95(3).

\bibitem[Nocedal, 1980]{Nocedal:1980}
Nocedal, J. (1980).
\newblock Updating quasi-newton matrices with limited storage.
\newblock {\em Mathematics of computation}, 35(151):773--782.

\bibitem[Patra et~al., 2005]{Patra:etal:2005}
Patra, A., Bauer, A., Nichita, C., Pitman, E., Sheridan, M., Bursik, M., Rupp,
  B., Webber, A., Stinton, A., Namikawa, L., and Renschler, C. (2005).
\newblock Parallel adaptive numerical simulation of dry avalanches over natural
  terrain.
\newblock {\em Journal of Volcanology and Geothermal Research}, 139(1):1--21.
\newblock Modeling and Simulation of Geophysical Mass Flows.

\bibitem[Pourahmadi, 1999]{Pourahmadi:1999}
Pourahmadi, M. (1999).
\newblock Joint mean-covariance models with applications to longitudinal data:
  Unconstrained parameterisation.
\newblock {\em Biometrika}, 86(3):677--690.

\bibitem[Pourahmadi, 2000]{Pourahmadi:2000}
Pourahmadi, M. (2000).
\newblock Maximum likelihood estimation of generalised linear models for
  multivariate normal covariance matrix.
\newblock {\em Biometrika}, 87(2):425--435.

\bibitem[Richards, 1931]{Richards:1931}
Richards, L.~A. (1931).
\newblock Capillary conduction of liquids through porous mediums.
\newblock {\em physics}, 1(5):318--333.

\bibitem[Sacks et~al., 1989]{Sack:Schi:Welc:1989}
Sacks, J., Schiller, S.~B., and Welch, W.~J. (1989).
\newblock Designs for computer experiments.
\newblock {\em Technometrics}, 31(1):41--47.

\bibitem[Salvucci and Entekhabi, 1994]{Salv:Ente:1994}
Salvucci, G.~D. and Entekhabi, D. (1994).
\newblock Equivalent steady soil moisture profile and the time compression
  approximation in water balance modeling.
\newblock {\em Water Resources Research}, 30(10):2737--2749.

\bibitem[Santner et~al., 2018]{Sant:Will:Notz:2018}
Santner, T.~J., Williams, B.~J., and Notz, W.~I. (2018).
\newblock {\em The Design and Analysis of Computer Experiments}.
\newblock Springer Series in Statistics. Springer-Verlag, New York, NY, second
  edition.

\bibitem[Santner et~al., 2003]{Sant:Will:2003}
Santner, T.~J., Williams, B.~J., Notz, W.~I., and Williams, B.~J. (2003).
\newblock {\em The design and analysis of computer experiments}, volume~1.
\newblock Springer.

\bibitem[Sch\"{a}fer et~al., 2021]{Schaf:Sulli:Owha:2021}
Sch\"{a}fer, F., Sullivan, T.~J., and Owhadi, H. (2021).
\newblock Compression, inversion, and approximate pca of dense kernel matrices
  at near-linear computational complexity.
\newblock {\em Multiscale Modeling \& Simulation}, 19(2):688--730.

\bibitem[Schäfer et~al., 2021]{Scha:Katz:Owha:2021}
Schäfer, F., Katzfuss, M., and Owhadi, H. (2021).
\newblock Sparse cholesky factorization by kullback-leibler minimization.

\bibitem[Spiller et~al., 2023]{Spill:etal:2023}
Spiller, E., Wolpert, R., Tierz, P., and Asher, T. (2023).
\newblock The zero problem: Gaussian process emulators for range-constrained
  computer models.
\newblock {\em SIAM/ASA Journal on Uncertainty Quantification}, 11:540--566.

\bibitem[Stein, 1999]{Stein:1999}
Stein, M.~L. (1999).
\newblock {\em Interpolation of spatial data: some theory for kriging}.
\newblock Springer Science \& Business Media.

\bibitem[Stein, 2002]{Stein:2002}
Stein, M.~L. (2002).
\newblock The screening effect in kriging.
\newblock {\em The Annals of Statistics}, 30(1):298 -- 323.

\bibitem[Tierz et~al., 2024]{Tierz:etal:2024}
Tierz, P., Spiller, E.~T., Clarke, B.~A., Dessalegn, F., Bekele, Y., Lewi, E.,
  Yirgu, G., Wolpert, R.~L., Loughlin, S.~C., and Calder, E.~S. (2024).
\newblock Topographic controls on pyroclastic density current hazard at aluto
  volcano (ethiopia) identified using a novel zero-censored gaussian process
  emulator.
\newblock {\em Journal of Geophysical Research: Solid Earth},
  129(5):e2023JB028645.
\newblock e2023JB028645 2023JB028645.

\bibitem[Vecchia, 1988]{Vecchia:1988}
Vecchia, A.~V. (1988).
\newblock Estimation and model identification for continuous spatial processes.
\newblock {\em Journal of the Royal Statistical Society. Series B
  (Methodological)}, 50(2):297--312.

\bibitem[Welch et~al., 1992]{Welch:1992}
Welch, W.~J., Buck, R.~J., Sacks, J., Wynn, H.~P., Mitchell, T.~J., and Morris,
  M.~D. (1992).
\newblock Screening, predicting, and computer experiments.
\newblock {\em Technometrics}, 34(1):15--25.

\bibitem[Wijaya et~al., 2023]{Wijaya:etal:2023}
Wijaya, O.~T., Yang, T.-H., Hsu, H.-M., and Gourbesville, P. (2023).
\newblock A rapid flood inundation model for urban flood analyses.
\newblock {\em MethodsX}, 10:102202.

\bibitem[Williams and Rasmussen, 2006]{Will:Rasm:Edwa:2006}
Williams, C.~K. and Rasmussen, C.~E. (2006).
\newblock {\em Gaussian processes for machine learning}, volume~2.
\newblock MIT press Cambridge, MA.

\bibitem[Ypma, 2014]{Ypma:2014}
Ypma, J. (2014).
\newblock Introduction to nloptr: an r interface to nlopt.
\newblock {\em R Package}, 2:1--15.

\end{thebibliography}

\end{document}